\documentclass[sigconf]{acmart}

\AtBeginDocument{%
  \providecommand\BibTeX{{%
    \normalfont B\kern-0.5em{\scshape i\kern-0.25em b}\kern-0.8em\TeX}}}
    
\usepackage{amsmath}
\usepackage{graphicx}
\usepackage{CJKutf8}
\usepackage{multirow}
\usepackage{adjustbox}
\usepackage{tabularx}
\usepackage{booktabs}
\usepackage{tablefootnote}
\usepackage{booktabs}
\usepackage{array}
\usepackage{balance} 
\usepackage{multirow}
\usepackage[normalem]{ulem}
\usepackage{color}
\definecolor{lightgray}{RGB}{215,215,215}
\usepackage{colortbl}  
\usepackage{xcolor}
\useunder{\uline}{\ul}{}
\usepackage{subfigure}
\usepackage{algorithm}  
\usepackage{algorithmicx}  
\usepackage[noend]{algpseudocode}  

\usepackage{amsmath}  
\usepackage{enumitem}
\usepackage{tabularx}
\usepackage[utf8]{inputenc}
\usepackage[english]{babel}
\usepackage{amsthm}
\usepackage{bm}

\newcommand{\eg}{\emph{e.g., }}

\newcommand{\cf}{\emph{cf. }}

\newlength\myindent
\floatname{algorithm}{Algorithm}

\copyrightyear{2024}
\acmYear{2024}
\setcopyright{acmlicensed}\acmConference[SIGIR '24]{Proceedings of the 47th International ACM SIGIR Conference on Research and Development in Information Retrieval}{July 14--18, 2024}{Washington, DC, USA}
\acmBooktitle{Proceedings of the 47th International ACM SIGIR Conference on Research and Development in Information Retrieval (SIGIR '24), July 14--18, 2024, Washington, DC, USA}
\acmDOI{10.1145/3626772.3657825}
\acmISBN{979-8-4007-0431-4/24/07}

\begin{document}

\title{Denoising Diffusion Recommender Model}

\author{Jujia Zhao}
\email{zhao.jujia.0913@gmail.com}
\affiliation{
\institution{Leiden University}
\city{Leiden}
\country{The Netherlands}
}
\author{Wenjie Wang}
\email{wenjiewang96@gmail.com}
\authornote{Corresponding author. This work is supported by the National Key Research and Development Program of China (2022YFB3104701) and the National Natural Science Foundation of China (62272437).}
\affiliation{
\institution{National University of Singapore}
\country{Singapore}
}
\author{Yiyan Xu}
\email{yiyanxu24@gmail.com}
\affiliation{
\institution{University of Science and Technology of China}
\city{Hefei}
\country{China}
}
\author{Teng Sun}
\email{stbestforever@gmail.com}
\affiliation{
\institution{Shandong University}
\city{Qingdao}
\country{China}
}
\author{Fuli Feng}
\email{fulifeng93@gmail.com}
\affiliation{
\institution{University of Science and Technology of China}
\city{Hefei}
\country{China}
}
\author{Tat-Seng Chua}
\email{dcscts@nus.edu.sg}
\affiliation{
\institution{National University of Singapore}
\country{Singapore}
}

\begin{abstract}

Recommender systems often grapple with noisy implicit feedback. Most studies alleviate the noise issues from data cleaning perspective such as data resampling and reweighting, but they are constrained by heuristic assumptions. 
Another denoising avenue is from model perspective, which proactively injects noises into user-item interactions and enhances the intrinsic denoising ability of models.
However, this kind of denoising process poses significant challenges to the recommender model's representation capacity to capture noise patterns. 

To address this issue, we propose Denoising Diffusion Recommender Model (DDRM), which leverages multi-step denoising process of diffusion models to robustify user and item embeddings from any recommender models. DDRM injects controlled Gaussian noises in the forward process and iteratively removes noises in the reverse denoising process, thereby improving embedding robustness against noisy feedback. 
To achieve this target, the key lies in offering appropriate guidance to steer the reverse denoising process and providing a proper starting point to start the forward-reverse process during inference.
In particular, we propose a dedicated denoising module that encodes collaborative information as denoising guidance. 
Besides, in the inference stage, DDRM utilizes the average embeddings of users’ historically liked items as the starting point rather than using pure noise since pure noise lacks personalization, which increases the difficulty of the denoising process.
Extensive experiments on three datasets with three representative backend recommender models demonstrate the effectiveness of DDRM. 


\vspace{-0.1cm}
\end{abstract}


\maketitle

\section{Introduction}
\label{sec:introduction}

\begin{figure}
\setlength{\abovecaptionskip}{0cm}
\setlength{\belowcaptionskip}{-0.6cm}
\centering
\includegraphics[scale=0.8]
{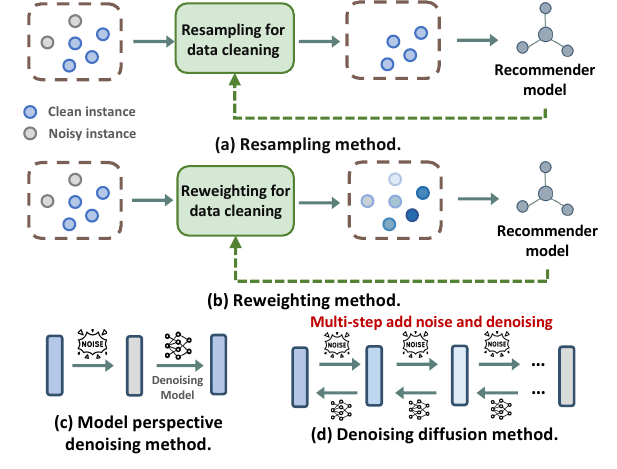}
\caption{Illustration of resampling, reweighting, model perspective denoising, and denoising diffusion methods.}
\label{fig:intro}
\end{figure}

Recommender systems play a pivotal role in personalized information delivery across a wide range of Web applications~\cite{yuan2023interaction,zhao2023popularity,guo2023towards}.    
Typically, recommender models learn personalized user preferences from user feedback~\cite{quan2023robust,zhang2023denoising}. 
Due to the ease of collecting implicit feedback (\eg click and purchase) in large volume, it has become indispensable for user preference learning~\cite{tian2022learning,zhang2023sled,xin2023improving}. 
Unfortunately, implicit feedback inevitably contains noises~\cite{chen2021autodebias,wang2021implicit,lin2023autodenoise}. 
For instance, clicks on micro-videos may not indicate users' actual satisfaction due to various interference factors~\cite{wang2021denoising}. 
Such noisy feedback misguides recommender models in interpreting user preferences, subsequently hampering the recommendation performance~\cite{wang2023efficient,fan2023graph,lin2023autodenoise}. 
As such, denoising implicit feedback for recommendation becomes an imperative task~\cite{chen2022denoising}.


Previous work primarily mitigates the impact of noisy feedback from the perspective of data cleaning, including resampling and reweighting user-item interactions. 
Specifically, 
1) resampling methods~\cite{ding2018improved,ding2019sampler,yu2020sampler} aim to identify noisy interactions and sample more clean interactions for training (Figure~\ref{fig:intro}(a)). 
For instance, WBPR~\cite{2011wbpr} believes that non-interacted popular items are more likely to be true negative items and allocates higher sampling probabilities. 
2) Reweighting methods~\cite{wang2023efficient,wang2021denoising,wang2022robust} utilize all training interactions yet assign lower weights to potential noisy ones (Figure~\ref{fig:intro}(b)). 
For instance, reweighted loss~\cite{wang2021denoising} assigns lower weights to the large-loss interactions since it assumes large-loss interactions are more likely to be noisy. 
Notably, these data cleaning methods depend on certain heuristic assumptions, such as the large-loss assumption~\cite{wang2021denoising} and cross-model agreement~\cite{wang2022robust}.
As their assumptions rely heavily on the distribution of noisy interactions, these data cleaning methods suffer from limited adaptability, requiring substantial configuration tuning to adapt to different backend models and datasets. 


For denoising implicit feedback, another research line is from model perspective, seeking to enhance recommender models' inherent noise resistance capabilities. 
These model perspective methods usually add random noises to user-item interactions~\cite{wu2016collaborative} or drop positive interactions as augmented data~\cite{wu2021sgl,gao2022self}, and then regulate recommender models to learn robust representations from the augmented data~\cite{wang2023robust}. 
For example, CDAE corrupts users' interactions randomly with a noise ratio and subsequently optimizes recommender models to recover the original clean interactions. 
Nevertheless, as illustrated in Figure~\ref{fig:intro}(c), these model perspective methods solely rely on a denoising model to directly convert noisy data into clean data,
imposing substantial demands on the model's representation capacity to efficiently capture noise patterns. 


Diffusion models, as a kind of powerful generative model, inherently possess a denoising aptitude to enhance existing model perspective denoising methods~\cite{ho2020denoising}. 
Diffusion models have already revealed remarkable representation capabilities across various domains like image generation and molecule generation~\cite{vignac2022digress,croitoru2022diffusion}. 
The potential benefits lie in two aspects: 
1) during the forward process, diffusion models enhance noise diversity by injecting noises with controllable noise scales and steps, potentially leading to the capability to denoise a broader spectrum of noise.
and 2) in the reverse denoising process, diffusion models decompose the complex denoising problem into multiple steps, thereby reducing the denoising difficulty at each step (Figure~\ref{fig:intro}(d)). 
In light of these, it is promising to leverage diffusion models to robustify user and item representations from existing recommender models.

To this end, we propose a plug-in denoising model for existing recommender models called \textbf{D}enoising \textbf{D}iffusion \textbf{R}ecommender \textbf{M}odel (DDRM). 
In the training stage, given user and item embeddings from any recommender models, DDRM improves their robustness against noisy feedback via a forward-reverse process. 
In the forward process, we proactively inject Gaussian noises into user and item embeddings with adjustable scales and steps, yielding noisy embeddings. 
The reverse denoising process then iteratively removes noises via a learnable neural network. 
During the inference phase, given a user, DDRM should choose a starting point to initialize this refined forward-reverse process, thereby generating an ideal item that aligns closely with the user's preferences.

To achieve the denoising target with DDRM, the key lies in 
1) offering appropriate guidance to steer the reverse denoising process and 
2) providing a proper starting point to start the forward-reverse process during inference. 
As for the guidance in the denoising process, our goal is to derive effective representations from weak collaborative information, which helps to guide the denoising process towards the clean embeddings of users (or items).
As for the starting point, we aim to incorporate personalized information rather than relying solely on pure noise. 
This strategy diverges from traditional diffusion models, which typically generate images from pure noise guided by abundant textual instructions~\cite{li2022diffusion}. 
In the recommendation scenario, guidance through collaborative information is not as explicit as textual instructions, making it challenging to generate personalized ideal items if we use pure noise as the starting point (see evidence in Table~\ref{tab:backbone}). 
By incorporating personalized information, it's more likely to generate the ideal item closely aligned with individual user preference.

Toward these goals, we design a denoising module for the reverse process of DDRM. 
Given noisy user (or item) embeddings, the denoising module devises strategies to encode the collaborative information, \eg users' liked items, to guide the reverse denoising process. 
For the inference phase, to generate an ideal item as the recommendation for a user, we take the average embeddings of the user's historically liked items as the starting point, instead of using pure noise (\cf Section~\ref{sec:inference} for details). 
Given the embedding of the generated item, we present a rounding function to ground the generated item to existing item candidates by the embeddings' similarity. 
As an extension, we also consider adding a reweighted loss to supplement DDRM from the perspective of data cleaning. 
We implement DDRM on three representative recommender models and conduct comprehensive experiments to validate its effectiveness against other baselines on three public datasets. 

The main contributions of this work are threefold:
\begin{itemize}[leftmargin=*]
    \item We propose a model-agnostic denoising diffusion recommender model, which robustifies the user and item representations from any recommender models against noisy feedback. 
    \item We design the user and item denoising modules to denoise the user and item embeddings, in which we incorporate collaborative information as guidance to steer the denoising process, and integrate personalized information as the starting point during inference.
    \item We instantiate DDRM on three backend models and execute extensive experiments under various settings, confirming its efficacy across three public datasets\footnote{Our code and data are released at \url{https://github.com/Polaris-JZ/DDRM}.}. 
\end{itemize}
\section{Preliminary }
\label{sec:preliminary}
Diffusion models have already demonstrated proficient performance in domains like computer vision and molecular generation~\cite{ruiz2023dreambooth,huang2023mdm}. 
Typically, diffusion models encompass two components: the forward and reverse processes~\cite{ho2020denoising}.

\begin{figure*}
\setlength{\abovecaptionskip}{0.03cm}
\setlength{\belowcaptionskip}{-0.2cm}
\centering
\includegraphics[scale=0.9]
{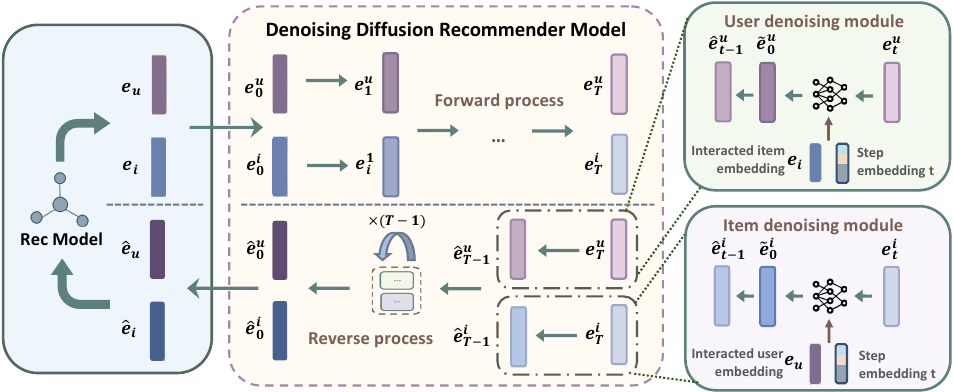}
\caption{Structure of DDRM. The left part is the backend  recommender model. DDRM accepts both user and item embeddings as inputs and subsequently produces denoised embeddings that are fed back into the model to do the recommendation task.}
\label{fig:method}
\end{figure*}

\noindent$\bullet$ \textbf{Forward process} aims to inject Gaussian noises into the original data. Given a data sample $\bm{x_0}$, diffusion models continuously add different scale Gaussian noises to it in $T$ steps until get $\bm{x_T}$. Specifically, for adding noise from $\bm{x_{t-1}}$ to $\bm{x_t}$, we have:
\begin{equation}
\label{eqn:forward1}
\begin{aligned}
& q(\bm{x_t}|\bm{x_{t-1}}) = \mathcal{N}(\bm{x_t},\sqrt{1-\beta_t}\bm{x_{t-1}},\beta_t \mathbf{I}), \\
\end{aligned}
\end{equation}
where $t\in\{{1,2,...,T}\}$ is the current step, $\beta_t \in (0,1)$ is the noise scale in step $t$, $\bm{I}$ is the identity matrix, and $\mathcal{N}$ is the Gaussian distribution which means $\bm{x_t}$ is sampled from this distribution. According to the additivity of independent Gaussian noises and reparameterization trick~\cite{ho2020denoising,hoogeboom2021argmax}, $\bm{x_t}$ can be directly obtained from $\bm{x_0}$ in the calculation:
\begin{equation}
\label{eqn:forward2}
\begin{aligned}
& q(\bm{x_t}|\bm{x_0}) = \mathcal{N}(\bm{x_t},\sqrt{\bar{\alpha_t}}\bm{x_{0}},(1-\bar{\alpha_t}) \mathbf{I}), \\
\end{aligned}
\end{equation}
where $\bar{\alpha_t}=\prod_{s=1}^t \alpha_s$, $\alpha_s=1-\beta_s$.

\noindent$\bullet$ \textbf{Reverse process} is designed to iteratively denoise the noisy data $\bm{x_T}$, following the sequence ($\bm{x_T}\rightarrow\bm{x_{T-1}} \rightarrow \bm{x_{T-2}} \rightarrow ... \rightarrow \bm{x_0}$).    
According to \cite{yang2022diffusion}, under the conditions where $q(\bm{x_t} | \bm{x_{t-1}})$ conforms to a Gaussian distribution and $\beta_t$ remains sufficiently small, the distribution  $p(\bm{x_{t-1}} | \bm{x_t})$ also exhibits Gaussian properties. 
As such, a neural network can be utilized to predict this reverse distribution:
\begin{equation}
\label{eqn:forward3}
\begin{aligned}
& p(\bm{x_{t-1}}|\bm{x_t})=\mathcal{N}(\bm{x_{t-1}};\bm{\mu_\theta}(\bm{x_t},t),\bm{\Sigma_\theta}(\bm{x_t},t)),
\\
\end{aligned}
\end{equation}
where $\bm{\theta}$ is the parameters of the neural network, and $\bm{\mu_\theta}(\bm{x_t},t)$ and $\bm{\Sigma_\theta}(\bm{x_t},t)$ are the mean and covariance of this Gaussian distribution.



\noindent$\bullet$ \textbf{Training.}
For training the diffusion models, the key focus is obtaining reliable values for $\bm{\mu}_\theta(\bm{x}_t,t)$ and $\bm{\Sigma}_\theta(\bm{x}_t,t))$ to guide the reverse process towards accurate denoising. To achieve this, it is important to optimize the variational lower bound of the negative log-likelihood of the model's predictive denoising distribution $p_\theta(\bm{x_0})$ :
\begin{equation}
\label{eq:train1}
\begin{aligned}
\mathcal{L} & = \mathbb{E}_{q(x_0)}\left[-\log p_\theta(\bm{x}_0) \right] \\
& \le 
\mathbb{E}_q\left[L_T + L_{T-1} + ... + L_0 \right], \rm where\\ 
\end{aligned} 
\end{equation}

\begin{equation}
\label{eq:train2}
\left\{
\begin{aligned}
& L_T = D_{\text{KL}}(q(\bm{x}_T|\bm{x}_0)\parallel p_\theta(\bm{x}_T))\\
& L_t = D_{\text{KL}}(q(\bm{x}_t|\bm{x}_{t+1},\bm{x}_0)\parallel p_\theta(\bm{x}_t|\bm{x}_{t+1}))  \\
& L_0 = -\log p_\theta(\bm{x}_0|\bm{x}_1),
\end{aligned}
\right.
\end{equation}
where $t\in\{{1,2,...,T-1}\}$. While $L_T$ can be disregarded during training due to the absence of learnable parameters in the forward process, $L_0$ represents the negative 
log probability of the original data sample $x_0$ given the first-step noisy data $x_1$,
and $L_t$ aims to align the distribution $p_\theta(\bm{x}_t|\bm{x}_{t+1})$ with the tractable posterior distribution $q(\bm{x}_t|\bm{x}_{t+1},\bm{x}_0)$ in the reverse process~\cite{luo2022understanding}.

\section{Method}
\label{sec:method}




To mitigate the effect of noisy implicit feedback, we propose DDRM to denoise the user and item embeddings. 
As illustrated in Figure~\ref{fig:method}, given pre-trained user and item embeddings, DDRM continuously injects Gaussian noises and then denoises these noisy embeddings iteratively through a forward-reverse process. 
For effective guidance during denoising, DDRM incorporates specialized user and item denoising modules within the reverse process.
During the inference phase, DDRM applies this refined forward-reverse process to a specifically chosen starting point, thereby generating an ideal item that aligns closely with user’s preferences.

\subsection{DDRM Framework}\label{sec:framework}
\vspace{3pt}
\noindent$\bullet$ \textbf{Forward process.} 
Given pre-trained user embeddings $\bm{e_u}$ of user $u$ and item embeddings $\bm{e_i}$ of item $i$ from a backend recommender model, 
we begin the forward process by setting $\bm{e_0^u}=\bm{e_u}$ and $\bm{e_0^i}=\bm{e_i}$. 
Subsequently, we continuously incorporate Gaussian noises into $\bm{e_0^u}$ and  $\bm{e_0^i}$ separately with adjustable scales and steps:
\begin{equation}
\label{eqn:f1}
\begin{aligned}
& q(\bm{e_t^u}|\bm{e_0^u}) = \mathcal{N}(\bm{e_t^u},\sqrt{\bar{\alpha_t}}\bm{e_0^{u}},(1-\bar{\alpha_t}) \mathbf{I}), \\
\end{aligned}
\end{equation}
\begin{equation}
\label{eqn:f2}
\begin{aligned}
& q(\bm{e_t^i}|\bm{e_0^i}) = \mathcal{N}(\bm{e_t^i},\sqrt{\bar{\alpha_t}}\bm{e_0^{i}},(1-\bar{\alpha_t}) \mathbf{I}), \\
\end{aligned}
\end{equation}
where $\bar{\alpha_t}=\prod_{s=1}^t \alpha_s$, $\alpha_s=1-\beta_s$, $\beta_s \in (0,1)$ controls the noise scale added to the embedding in the current step $s$, and $e_{t}^{(\cdot)}$ denotes the user or item embeddings in the forward step $t$. 
To regulate the noise level in each step, we follow \cite{diffrec} employing a \textit{linear variance} noise schedule in the forward process:
\begin{equation}
    1 - \bar{\alpha}_t = s\cdot\left[\alpha_{\min} + \dfrac{t-1}{T-1}(\alpha_{\max} - \alpha_{\min})\right],
\label{eq:f3}
\end{equation}
where $\alpha_{min}$ and $\alpha_{max}$ are the minimum and maximum of the noise correspondingly, $t$ is the current forward step, $T$ is the total forward step, and $s \in (0,1)$ controls the noise scale.

\vspace{3pt}
\noindent$\bullet$ \textbf{Reverse process.} 
After getting noisy user embeddings $\bm{e_T^u}$ and noisy item embeddings $\bm{e_T^i}$ in the forward process, we denoise these embeddings iteratively in the reverse process. 
In each reverse step, we design the user denoising module and the item denoising module to denoise user embeddings and item embeddings separately since the noise distribution is different for users and items:
\begin{equation}
\label{eqn:r1}
\begin{aligned}
& p_\theta(\bm{\hat{e}_{t-1}^{u}}|\bm{\hat{e}_t^u})=\mathcal{N}(\bm{\hat{e}_{t-1}^{u}};\bm{\mu_\theta}(\bm{\hat{e}_t^u}, t),\bm{\Sigma_\theta}(\bm{\hat{e}_t^u}, t)),
\\
\end{aligned}
\end{equation}
\begin{equation}
\label{eqn:r2}
\begin{aligned}
& p_\psi(\bm{\hat{e}_{t-1}^{i}}|\bm{\hat{e}_t^i})=\mathcal{N}(\bm{\hat{e}_{t-1}^{i}};\bm{\mu_\psi}(\bm{\hat{e}_t^i}, t),\bm{\Sigma_\psi}(\bm{\hat{e}_t^i}, t)),
\\
\end{aligned}
\end{equation}
where $\bm{\hat{e}_{t}^{u}}$ and $\bm{\hat{e}_{t}^{i}}$ are the denoised embeddings in the reverse step $t$,
$\bm{\theta}$ and $\bm{\psi}$ are the learnable parameters of the user denoising module and the item denoising module correspondingly. 
These denoising module are executed iteratively in the reverse process until the generation of final clean embeddings $\bm{
\hat{e}_0^u}$ and $\bm{\hat{e}_0^i}$.

\vspace{3pt}
\noindent$\bullet$ \textbf{Denoising module.} 
Denoising module aims to denoise the noisy embedding in each reverse step.
Since the user denoising module and the item denoising module have the same structure, we mainly focus on explaining the user denoising module as formulated as Eq. (\ref{eqn:r1}). 
As illustrated in Eq. (\ref{eq:train2}), the diffusion training aims to align the distribution $p_\theta(\bm{\hat{e}_{t-1}^{u}}|\bm{\hat{e}_t^u})$ with the tractable posterior distribution $q(\bm{e_{t-1}^u}|\bm{e_t^u},\bm{e_0^u})$ in the reverse process, thus we can use $q(\bm{e_{t-1}^u}|\bm{e_t^u},\bm{e_0^u})$ to constrain $p_\theta(\bm{\hat{e}_{t-1}^{u}}|\bm{\hat{e}_t^u})$.
Through Bayes' theorem, we can derive:
\begin{equation}
\small
\label{eqn:d1}
\begin{aligned}
& q(\bm{e_{t-1}^u}|\bm{e_t^u},\bm{e_0^u}) = \mathcal{N}(\bm{e_{t-1}^u}, \tilde{\bm{\mu_t}}(\bm{e_t^u},\bm{e_0^u}),\tilde\beta_t \bm{I}), \quad\rm where\\ 
\end{aligned}
\end{equation} 
\begin{equation}
\small
\label{eq:d2}
\left\{
\begin{aligned}
&\tilde{\bm{\mu_t}}(\bm{e_t^u},\bm{e_0^u}) =\dfrac{\sqrt{\alpha_t}(1-\bar{\alpha}_{t-1})}{1-\bar{\alpha}_t}\bm{e_t^u}+\dfrac{\sqrt{\bar{\alpha}_{t-1}}(1-\alpha_t)}{1-\bar{\alpha}_t}\bm{e_0^u}, \\
&\tilde \beta_t =\dfrac{(1-\alpha_t)(1-\bar{\alpha}_{t-1})}{1-\bar{\alpha}_t}.
\end{aligned}
\right.
\end{equation}
$\tilde{\bm{\mu_t}}(\bm{e_t^u},\bm{e_0^u})$ and $\tilde\beta_t \bm{I}$ are the mean and covariance of $q(\bm{e_{t-1}^u}|\bm{e_t^u},\bm{e_0^u})$. 
Following~\cite{diffrec}, we can similarity factorize $p_\theta(\bm{\hat{e}_{t-1}^{u}}|\bm{\hat{e}_t^u})$:
\begin{equation}
\small
\label{eqn:d3}
\begin{aligned}
& p_\theta(\bm{\hat{e}_{t-1}^{u}}|\bm{\hat{e}_t^u})=\mathcal{N}(\bm{\hat{e}_{t-1}^{u}};\bm{\mu}_\theta(\bm{\hat{e}_t^u}, t),\tilde\beta_t \bm{I})), \quad\rm where
\end{aligned}
\end{equation}
\begin{equation}
\small
\label{eq:d4}
\begin{aligned}
&{\bm{\mu_\theta}}(\bm{\hat{e}_t^u}, t) =\dfrac{\sqrt{\alpha_t}(1-\bar{\alpha}_{t-1})}{1-\bar{\alpha}_t}\bm{\hat{e}_t^u}+\dfrac{\sqrt{\bar{\alpha}_{t-1}}(1-\alpha_t)}{1-\bar{\alpha}_t}\bm{\tilde{e}_0^u}.
\end{aligned}
\end{equation}
$\bm{\tilde{e}_0^u}$ is the predicted $\bm{e_0^u}$ since the distribution of $\bm{e_0^u}$ is unknown in the reverse process. We employ the multi-layer perceptron (MLP) to reconstruct $\bm{e_0^u}$ in the denoising module. 

To ensure tractability in the embedding denoising process, it is paramount to utilize specific conditions as guidance.
In this context, DDRM incorporates two key conditions during the denoising phase: collaborative information and step information. 
Collaborative information, which is based on user interaction behaviors, can enable the denoising module to recognize user preferences, subsequently assisting to identify and mitigate noise. 
Additionally, the step information also affects the denoising performance.
It provides insight into the current noise level in the embeddings, offering a gauge on the extent of denoising required at each step.
As such, we use collaborative information $\bm{c_u}$ and step embedding as condition elements, to guide the reconstruction process in the denoising modules (\cf Section~\ref{sec:optimization} for the calculation of $\bm{c_u}$ in detail). 
Specifically, for user embedding $\bm{\hat{e}_t^u}$ in the reverse step $t$, the user reconstruction MLP yields:
\begin{equation}
\label{eqn:d5}
\begin{aligned}
& \bm{\tilde{e}_0^u}={\bm{f_\theta}}(\bm{\hat{e}_t^u}, \bm{c_u}, t), \\ 
\end{aligned}
\end{equation} 
where 
$\bm{\tilde{e}_0^u}$ is the predicted $\bm{e_0^u}$ by the reconstruction MLP with the parameter $\bm{\theta}$. 
It is noteworthy that the step information is encoded through sinusoidal positional encoding~\cite{ho2020denoising}, and these three inputs are concatenated together to feed into the MLP.

Similarly, for the item denoising module, given item embedding $\bm{\hat{e}_t^i}$ in the reverse step $t$, the other item reconstruction MLP outputs $\bm{\tilde{e}_0^i}={\bm{f_\psi}}(\bm{\hat{e}_t^i}, \bm{c_i}, t)$, where $\bm{c_i}$ represent the collaborative information of item $i$,
and $\bm{\tilde{e}_0^i}$ is the predicted $\bm{e_0^i}$ by the item reconstruction MLP with the parameter $\bm{\psi}$.

Generally, in the denoising module, we get predicted original embedding $\bm{\tilde{e}_0^u}$ and $\bm{\tilde{e}_0^i}$ from the user reconstruction MLP and the item reconstruction MLP, respectively, and then utilize Eq. (\ref{eqn:d3})  to get the denoised embeddings for the current step. 

\begin{algorithm}[t]
	\caption{\textbf{DDRM Training}}  
	\label{algo:training}
	\begin{algorithmic}[1]
		\Require interaction data $\bm{\bar{D}}$, pre-trained user embedding $\bm{e_0^u}$, pre-trained item embedding $\bm{e_0^i}$, diffusion step $T$, user reconstruction MLP $\bm{f_\theta}$, item reconstruction MLP $\bm{f_\psi}$
            \Repeat 
            \State Sample a batch of interactions $\bm{D} \subset \bm{\bar{D}}$.
            \ForAll{$(u,i,j) \in \bm{D}$}
            \State Sample $t\sim\mathcal{U}(1,T)$
            \State Compute $\bm{e}_t^u$ given $\bm{e}_0^u$ and $t$ via $q(\bm{e}_t^u|\bm{e}_0^u)$ in Eq. (\ref{eqn:f1});
            \State Compute $\bm{e}_t^i$ given $\bm{e}_0^i$ and $t$ via $q(\bm{e}_t^i|\bm{e}_0^i)$ in Eq. (\ref{eqn:f2});
            \State Reconstruct $\bm{\tilde{e}_0^u}$ and $\bm{\tilde{e}_0^i}$ through $\bm{f_\theta}$ and $\bm{f_\psi}$;
            \State Calculate $\mathcal{L}_{\text{final}}$ by Eq. (\ref{eq:l4});
            \State Take gradient descent step on $\nabla_\theta(\mathcal{L}_{\text{final}})$ to optimize $\theta$;                
            \State Take gradient descent step on$\nabla_\psi(\mathcal{L}_\text{final})$ to optimize $\psi$;
            \EndFor
            \Until{converged}
            \Ensure optimized $\theta$, $\psi$.
	\end{algorithmic}
\end{algorithm}
\setlength{\textfloatsep}{0.28cm}

\subsection{Optimization}\label{sec:optimization}
\vspace{3pt}
\noindent$\bullet$ \textbf{DDRM training.} 
The optimization of DDRM is under the BPR training setting: 
given recommendation data with the triplet $(u,i,j)$, item $i$ and item $j$ are the positive item and negative item of user $u$, respectively.
Please note that we only conduct denoising for user $u$ and positive interacted item $i$ 
since denoising negative item $j$ is rendered relatively insignificant due to the random negative sampling mechanism.

To optimize the embedding denoising process,
it is essential to minimize the variational lower bound of the predicted user and item embeddings. 
According to the KL divergence based on the multivariate Gaussian distribution in Eq. (\ref{eq:train2}), the
reconstruction loss of the denoising process within a training iteration is expressed as:
\begin{equation}
\small
\label{eq:o1}
\begin{aligned}
\mathcal{L}_{\text{re}}(u,i) & = \mathbb{E}_q\left[-\log p_{\bm{\theta}}(\bm{\hat{e}}_0^u) -\log p_{\bm{\psi}}(\bm{\hat{e}}_0^i) \right] \\
& \le 
\mathcal{L}^u + \mathcal{L}^i + \mathcal{L}_0^u + \mathcal{L}_0^i, \quad \rm where\\ 
\end{aligned} 
\end{equation}
\begin{equation}
\small
\label{eq:o2}
\left\{
\begin{aligned}
\mathcal{L}^u 
&=
\textstyle\sum_{t=2}^{T}\mathbb{E}_{q}\left[\dfrac{1}{2}\left(\dfrac{\bar{\alpha}_{t-1}}{1-\bar{\alpha}_{t-1}}-\dfrac{\bar{\alpha}_{t}}{1-\bar{\alpha}_{t}}\right)||\bm{e_0^u}-\hat{\bm{f_\theta}}(\bm{e_t^u}, \bm{e_{i}}, t)||^2_2\right],  \\
\mathcal{L}^i 
&=
\textstyle\sum_{t=2}^{T}\mathbb{E}_{q}\left[\dfrac{1}{2}\left(\dfrac{\bar{\alpha}_{t-1}}{1-\bar{\alpha}_{t-1}}-\dfrac{\bar{\alpha}_{t}}{1-\bar{\alpha}_{t}}\right)||\bm{e_0^i}-\hat{\bm{f_\psi}}(\bm{e_t^i}, \bm{e_{u}}, t)||^2_2\right], \\
\mathcal{L}_0^u
&=
\mathbb{E}_{q}\left[\parallel \bm{e_0^u} - \hat{\bm{f_\theta}}(\bm{e_1^u}, \bm{e_{i}},1)\parallel_2^2\right], \\
\mathcal{L}_0^i
&=
\mathbb{E}_{q}\left[\parallel \bm{e_0^i} - \hat{\bm{f_\theta}}(\bm{e_1^i},  \bm{e_{u}},1)\parallel_2^2\right], \\
\end{aligned} 
\right.
\end{equation}
where $\bm{e_i}$ and $\bm{e_u}$ are the original embeddings of user $u$ and item $i$, which serve as the collaborative information $\bm{c_u}$ and $\bm{c_i}$, respectively. 
$\mathcal{L}^u$  and $\mathcal{L}^i$ are the user and item reconstruction loss in the reverse process, $\mathcal{L}_0^u$ and $\mathcal{L}_0^i$ are the final prediction loss correspondingly.
From Eq. (\ref{eq:o3}), it is clear that the essence of the DDRM training lies in optimizing the distance between the reconstructed embedding derived from MLP and the original embedding.

To reduce the computational cost in the implementation, we simplify Eq. (\ref{eq:o2}) by uniformly sampling $t$ from $\{{1,2,...,T}\}$ instead of summing $T$ steps and removing the weight before the MSE terms:

\begin{equation}
\small
\label{eq:o3}
\begin{aligned}
\mathcal{L}_{\text{re}}(u,i)
=
(\mathcal{L}_{\text{simple}}^u + \mathcal{L}_{\text{simple}}^i)/2, \quad \rm where\\ 
\end{aligned} 
\end{equation}
\begin{equation}
\small
\label{eq:o4}
\left\{
\begin{aligned}
\mathcal{L}_{\text{simple}}^u
&=
\mathbb{E}_{t\sim \mathcal{U}(1,T)}\mathbb{E}_{q}\left[||\bm{e_0^u}-\hat{\bm{e_\theta}}(\bm{e_t^u}, \bm{e_{i}}, t)||^2_2\right], \\
\mathcal{L}_{\text{simple}}^i 
&=
\mathbb{E}_{t\sim \mathcal{U}(1,T)}\mathbb{E}_{q}\left[||\bm{e_0^i}-\hat{\bm{e_\psi}}(\bm{e_t^i}, \bm{e_{u}}, t)||^2_2\right]. \\
\end{aligned} 
\right.
\end{equation}

\vspace{3pt}
\noindent$\bullet$ \textbf{Loss function.} 
The final loss function of DDRM comprises two parts: a reconstruction loss for the denoising process and a BPR loss for the recommendation task. 
The reconstruction loss $\mathcal{L}_{\text{re}}(u,i)$ is derived from Eq. (\ref{eq:o3}), which regulates the denoising of user and item embeddings.

After obtaining the denoised user and positive item embeddings via DDRM, these embeddings contribute to the computation of BPR loss $\mathcal{L}_{\text{bpr}}$~\cite{he2020lightgcn}. We design a loss balance factor $\lambda$ to adjust the weight of these two losses:
\begin{equation}\label{eq:l2}
\begin{aligned}
\mathcal{L}(u,i,j) = \lambda \mathcal{L}_{\text{bpr}}(u,i,j) + (1-\lambda)\mathcal{L}_{\text{re}}(u,i), 
\end{aligned} 
\end{equation}
where $i$ and $j$ are positive and negative items for user $u$ in the BPR training setting. 
As an extension, we also consider adding a reweighted loss to supplement DDRM from the perspective of data cleaning (see empirical evidence of its effectiveness in Section~\ref{sec:ablation}).
Specifically, inspired by~\cite{wang2021denoising}, we dynamically allocate lower weights to instances with relatively lower positive scores since they are more likely to be noisy data.
\begin{equation}\label{eq:l3}
\begin{aligned}
w(u,i,j) = \text{sigmoid}(s(u,i))^\gamma, 
\end{aligned} 
\end{equation}
\begin{equation}\label{eq:l4}
\begin{aligned}
\mathcal{L}_{\text{final}}(u,i,j) = w(u,i,j)\mathcal{L}(u,i,j), 
\end{aligned} 
\end{equation}
where $s(u,i)$ quantifies the score between users $u$ and positive items $i$, and $\gamma$ is the reweighted factor which controls the range of weights.
The training step of DDRM is illustrated in Algorithm~\ref{algo:training}. 

\begin{algorithm}[t]
	\caption{\textbf{DDRM Inference}}  
	\label{algo:infer}
	\begin{algorithmic}[1]
		\Require all users $\bm{\bar{U}}$, diffusion step $T$, item reconstruction MLP $\bm{f_\psi}$
            \State Sample a batch of users $\bm{U} \subset \bm{\bar{U}}$.
            \ForAll{$u \in \bm{U}$}
            \State Compute average embeddings of users' historically liked items $\bm{\bar{e}_i}$ via Eq. (\ref{eq:i1});
            \State Compute $\bm{\bar{e}_T^i}$ given $\bm{\bar{e}_0^i}$ and $T$ via $q(\bm{\bar{e}_T^i}|\bm{\bar{e}_0^i})$ in Eq. (\ref{eqn:f2});
            \For{$t=T,\dots,1$}
            \State Reconstruct $\bm{\hat{e}_0^i}$ through $\bm{f_\psi}$;
            \State Compute $\bm{\hat{e}_{t-1}^i}$ from $\bm{\hat{e}_{t}^i}$ and $\bm{\hat{e}_0^i}$ via Eq. (\ref{eqn:d3});
            \EndFor
            \State Rounding via Eq. (\ref{eq:i2}) to get the ideal item embedding $\bm{e}$;
            \EndFor
            \Ensure the ideal item embedding $\bm{e}$.
    \end{algorithmic}
\end{algorithm}
\setlength{\textfloatsep}{0.28cm}

\subsection{Inference}\label{sec:inference}
In the inference phase, we need to apply the refined forward-reverse process to a starting point, thereby generating ideal items to do the recommendation task for each user.
Since guidance through collaborative information is not as explicit as textual instructions used in fields like image generation, we incorporate personalized information into the starting point rather than using pure noise.
Particularly, we take the average embeddings of users' historically liked items as input since the interaction information reflects the preferences of users, thereby it’s more likely to align the generated ideal item closely with individual user needs.
Specifically, we first get the average item embedding $\bm{\bar{e}_i}$:
\begin{equation}\label{eq:i1}
\begin{aligned}
\bm{\bar{e}_i} = \frac{1}{n}\textstyle\sum\limits_{i\in \mathcal{I}_u}\bm{e_i}, 
\end{aligned} 
\end{equation}
where $n$ is the number of historical interacted items of user $u$, and $\mathcal{I}_u$ is the historical interacted items.
Subsequently, we introduce noise into $\bm{\bar{e}_i}$ continuously following the sequence $\bm{\bar{e}_0^i}\rightarrow\bm{\bar{e}_1^i}  \rightarrow \cdots \rightarrow \bm{\bar{e}_T^i}$ in the forward process.    
And then, we set $\bm{\hat{e}_{T}^i}=\bm{\bar{e}_T^i}$ to execute the reverse process by $\bm{\hat{e}_{T}^i}\rightarrow\bm{\hat{e}_{T-1}^i}  \rightarrow \cdots \rightarrow \bm{\hat{e}_0^i}$ to generate a new item embedding $\bm{\hat{e}_0^i}$ conditioned on current step embedding and user original embedding $\bm{e_u}$. 
Following this, to obtain ideal items for the recommendation task, we develop a rounding function $s(\bm{\hat{e}_0^i},\bm{e_i})$ that calculates the inner product between the generated item embeddings $\bm{\hat{e}_0^i}$ and the candidate item embeddings to get a similarity score:
\begin{equation}\label{eq:i2}
\begin{aligned}
s(\bm{\hat{e}_0^i},\bm{e_i}) = \bm{\hat{e}_0^i} \cdot {\bm{e_i}}, 
\quad
i \in \mathcal{I}
\end{aligned} 
\end{equation}
where $\mathcal{I}$ is the candidate item pool.
Subsequently, we rank the similarity score and select the top-k candidate items for recommendation.
The inference procedure of DDRM is stated in Algorithm~\ref{algo:infer}. 

\section{Experiments}
\label{sec:experiment}

In this section, we conduct a comprehensive experimental study to address the following research questions:
\begin{enumerate}[leftmargin=*]
    \item [-] \textbf{RQ1:} How does the performance of DDRM compare with other baselines across the datasets in different experiment settings?
    \item [-] \textbf{RQ2:} What is the impact of different components (\eg reconstruction loss and reweighted loss, user and item denoising modules) within the DDRM on overall performance?
    \item [-] \textbf{RQ3:} How do design variations in DDRM influence efficacy?
\end{enumerate}

 \begin{table}[t]
\setlength{\belowcaptionskip}{0.1cm}
\caption{Statistics of three datasets under two distinct settings. ``\#Int.'' denotes interactions numbers. ``N'' and ``R'' represent natural noise setting and random noise setting, respectively.}
\label{tab:datasets_statistics}
\setlength{\tabcolsep}{1.2mm}{
\resizebox{0.48\textwidth}{!}{
\begin{tabular}{@{}lrrrrr@{}}
\hline
& \multicolumn{1}{l}{\textbf{\#User}} & \multicolumn{1}{l}{\textbf{\#Item (N)}} & \multicolumn{1}{l}{\textbf{\#Int. (N)}} & \multicolumn{1}{l}{\textbf{\#Item (R)}} & \multicolumn{1}{l}{\textbf{\#Int. (R)}} \\ \hline
\textbf{Yelp} & 54,574 & 77,405 & 1,471,675 & 34,395 & 1,402,736 \\
\textbf{Amazon-book} & 108,822 & 178,181 & 3,145,223 & 94,949 & 3,146,256 \\
\textbf{ML-1M} & 5,949 & 3,494 & 618,297 & 2,810 & 571,531 \\ \hline
\end{tabular}
}}
\end{table}
\begin{table*}[]
\setlength{\abovecaptionskip}{0cm}
\setlength{\belowcaptionskip}{0cm}
\caption{Overall performance of DDRM and other baselines under natural noise setting. Bold signifies the best performance among the backend models, model-agnostic methods and DDRM. * denotes statistically significant improvements of DDRM over the backend models, according to the t-tests with a significance level of $p$ < 0.01.}
\setlength{\tabcolsep}{3mm}{
\resizebox{\textwidth}{!}{
\label{tab:main_exp}
\begin{tabular}{l|cccc|cccc|cccc}
\hline
\multicolumn{1}{l|}{} & \multicolumn{4}{c|}{\textbf{Yelp}}                             & \multicolumn{4}{c|}{\textbf{Amazon-book}}                           & \multicolumn{4}{c}{\textbf{ML-1M}}                             \\
\textbf{Methods}      & \textbf{R@10}   & \textbf{R@20}   & \textbf{N@10}   & \textbf{N@20}   & \textbf{R@10}   & \textbf{R@20}   & \textbf{N@10}   & \textbf{N@20}   & \textbf{R@10}   & \textbf{R@20}   & \textbf{N@10}   & \textbf{N@20}   \\ \hline
\textbf{AdaGCL}       & 0.0464          & 0.0774          & 0.0277          & 0.0371          & 0.0282          & 0.0464          & 0.0166          & 0.0220& 0.0630& 0.1200& 0.0453          & 0.0659          \\
 \textbf{CDAE}         & 0.0305          & 0.0530& 0.0178          & 0.0246          & 0.0219          & 0.0399          & 0.0122          & 0.0175          & 0.0355          & 0.0675          & 0.0272          &0.0391          \\
\textbf{MultiVAE}     & 0.0484          & 0.0823          & 0.0289& 0.0391          & 0.0508          & 0.0771          & 0.0300& 0.0379          & 0.0636          & 0.1229          & 0.0450& 0.0667          \\
\textbf{DiffRec}      & 0.0501          & 0.0847          & 0.0307          & 0.0412          & 0.0537         & 0.0806          & 0.0329          & 0.0411          & 0.0658          & 0.1236          & 0.0488          & 0.0703          \\ \hline
\textbf{MFBPR}        & 0.0286          & 0.0503          & 0.0176          & 0.0242          & 0.0217          & 0.0379          & 0.0131          & 0.0179          & 0.0445          & 0.0890& 0.0429          & 0.0582          \\
\textbf{$\quad$+T-CE}         & 0.0316          & 0.0547          & 0.0191          & 0.0261          & 0.0230& 0.0393          & 0.0136          & 0.0185          & 0.0460& 0.0866& 0.0432& 0.0573\\
\textbf{$\quad$+R-CE}         & 0.0324          & 0.0554          & 0.0195          & 0.0265          & 0.0227          & 0.0394          & 0.0134          & 0.0184          & 0.0472& 0.0901          & 0.0434          & 0.0587          \\
\textbf{$\quad$+DeCA}         & 0.0295          & 0.0494          & 0.0181          & 0.0241          & 0.0118          & 0.0188          & 0.0071          & 0.0092          & 0.0451& 0.0863& 0.0428& 0.0851\\
\textbf{$\quad$+BOD}         & 0.0331& 0.0568& 0.0201& 0.0269& 0.0225& 0.0376& 0.0138& 0.0185& 0.0458& 0.0892& 0.0431& 0.0585\\  
\rowcolor[HTML]{D9D9D9} 
\textbf{$\quad$+DDRM}         & \textbf{0.0358*}& \textbf{0.0582*}& \textbf{0.0217*}& \textbf{0.0285*}& \textbf{0.0249*} & \textbf{0.0406*} & \textbf{0.0148*} & \textbf{0.0196*} & \textbf{0.0477*}& \textbf{0.0916*}& \textbf{0.0450*}& \textbf{0.0601*}\\ \hline
\textbf{LightGCN}     & 0.0502          & 0.0858          & 0.0295          & 0.0403          & 0.0432          & 0.0710& 0.0251          & 0.0333          & 0.0618          & 0.1193          & 0.0444          & 0.0652          \\
\textbf{$\quad$+T-CE}         & 0.0504          & 0.0856          & 0.0294          & 0.0400& 0.0421          & 0.0691          & 0.0242          & 0.0323          & 0.0625          & 0.1191          & 0.0457          & 0.0661          \\
\textbf{$\quad$+R-CE}         & \textbf{0.0516}& \textbf{0.0877}& 0.0304          & \textbf{0.0412}& 0.0439          & 0.0723          & 0.0253          & 0.0337          & 0.0623          & 0.1208          & 0.0457          & 0.0668          \\
\textbf{$\quad$+DeCA}         & 0.0486          & 0.0832          & 0.0286          & 0.0390& 0.0419          & 0.0688          & 0.0242          & 0.0321          & 0.0616          & 0.1202          & 0.0446          & 0.0659          \\
\textbf{$\quad$+BOD}         & 0.0481& 0.0821& 0.0278& 0.038& 0.0388& 0.0639& 0.0225& 0.0299& 0.0647& 0.1212& 0.0455& 0.0664\\ 
\rowcolor[HTML]{D9D9D9} 
\textbf{$\quad$+DDRM}         & \textbf{0.0516*} & 0.0870*& \textbf{0.0305*} & \textbf{0.0412*} & \textbf{0.0468*} & \textbf{0.0742*} & \textbf{0.0273*} & \textbf{0.0355*} & \textbf{0.0667*} & \textbf{0.1221*} & \textbf{0.0508*} & \textbf{0.0710*}\\ \hline
\textbf{SGL}          & 0.0485          & 0.0835          & 0.0287          & 0.0393          & 0.0467          & 0.0758          & 0.0267          & 0.0353          & 0.0620& 0.1164          & 0.0448          & 0.0648          \\
\textbf{$\quad$+T-CE}         & 0.0493          & 0.0840& 0.0293          & 0.0398          & 0.0483          & 0.0765          & 0.0276          & 0.0361          & 0.0647          & 0.1184          & 0.0470& 0.0667          \\
\textbf{$\quad$+R-CE}         & 0.0488          & 0.0831          & 0.0289          & 0.0393          & 0.0498          & 0.0772          & 0.0283          & 0.0367          & 0.0651          & 0.1165          & 0.0479          & 0.0670\\
\textbf{$\quad$+DeCA}         & 0.0476          & 0.0801          & 0.0282          & 0.0380& 0.0489          & 0.0764          & 0.0285          & 0.0368          & 0.0641          & 0.1183          & 0.0475          & 0.0673          \\
\textbf{$\quad$+BOD}         & 0.0505& 0.0838& 0.0303& 0.0403& 0.0517& 0.0801& 0.0300& 0.0385& 0.065& 0.1234& 0.0458& 0.0668\\ 
\rowcolor[HTML]{D9D9D9} 
\textbf{$\quad$+DDRM}         & \textbf{0.0517*} & \textbf{0.0860*}& \textbf{0.0312*} & \textbf{0.0415*} & \textbf{0.0535*}& \textbf{0.0813*}& \textbf{0.0313*}& \textbf{0.0396*}& \textbf{0.0698*} & \textbf{0.1261*} & \textbf{0.0530*}& \textbf{0.0739*} \\ \hline
\end{tabular}
}}
\end{table*}

\subsection{Experimental Settings}
\label{sec:setting}

\subsubsection{\textbf{Datasets.}} We evaluate our proposed DDRM on three publicly accessible datasets in different experiment settings. 1) \textbf{Yelp}\footnote{\url{https://www.yelp.com/dataset/.}} contains a large collection of user reviews and ratings for different restaurants. 2) \textbf{Amazon-book}\footnote{\url{https://jmcauley.ucsd.edu/data/amazon/.}} covers users' purchase history and rating scores over books. 3) \textbf{ML-1M}\footnote{\url{https://grouplens.org/datasets/movielens/1m/.}} compiles movie ratings submitted by users.
For each dataset, the interactions with ratings < 4 are regarded as false-positive interactions.

Following~\cite{diffrec}, we first arrange the user-item interactions chronologically based on the timestamps, and then split true-positive interactions (ratings $\ge$ 4) into training, validation and testing sets with a ratio of 7:1:2. 
To evaluate the effectiveness of denoising implicit feedback, we train and validate the framework on noisy interactions (both true-positive and false-positive interactions), and test the framework only on true-positive interactions. 
Specifically, we explore two types of noisy settings: natural noise and random noise. While keeping the testing set containing only true-positive interactions, 1) \textbf{Natural noise setting} introduces false-positive interactions (ratings < 4) into the original training and validation sets; 2) \textbf{Random noise setting} randomly samples unobserved interactions into the original training and validation sets. Moreover, we ensure that the training and validation sets under the two noisy settings are at the same scale as the original dataset partition. The statistics of datasets are shown in Table~\ref{tab:datasets_statistics}.

\subsubsection{\textbf{Baselines.}} To demonstrate the efficacy of our proposed DDRM in denoising implicit feedback, we compare DDRM with the state-of-the-art model-agnostic denoising methods. In particular,
1) \textbf{R-CE}\footnote{\url{https://github.com/WenjieWWJ/DenoisingRec}}~\cite{wang2021denoising} assumes large-loss interactions are more likely to be noisy and allocates lower weights to them.
2) \textbf{T-CE}~\cite{wang2021denoising}, guided by the same assumption as R-CE, directly eliminates large-loss interactions using a dynamic threshold.
3) \textbf{DeCA}\footnote{\url{https://github.com/wangyu-ustc/DeCA}}~\cite{wang2022robust} leverages predictions from different models as denoising signals, under the assumption that different models give more consistent predictions for clean data than for noisy data.
4) \textbf{BOD}\footnote{\url{https://github.com/CoderWZW/BOD}}~\cite{wang2023efficient} employs autoencoder-based generator to learn instance weight, thus denoising the data from data cleaning perspective.

Furthermore, we also compare DDRM with other competitive baselines including model perspective denoising methods and generative methods:
5) \textbf{AdaGCL}\footnote{\url{https://github.com/HKUDS/AdaGCL}}~\cite{jiang2023adaptive} is a graph collaborative filtering-based denoising method.
6) \textbf{CDAE}\footnote{\url{https://github.com/henry0312/CDAE}}~\cite{wu2016collaborative} introduces random noises to users' interactions during training and employs an auto-encoder for denoising.
7) \textbf{MultiVAE}\footnote{\url{https://github.com/dawenl/vae_cf}}~\cite{liang2018variational} utilizes variational auto-encoders with multinomial likelihood to model implicit feedback.
8) \textbf{DiffRec}\footnote{\url{https://github.com/YiyanXu/DiffRec}}~\cite{diffrec} is a diffusion-based generative recommender model that infers users' preferences by modeling the interaction probabilities in a denoising manner.

We implement DDRM and the aforementioned model-agnostic baselines to three representative backend models.
9) \textbf{MFBPR}\footnote{\url{https://github.com/guoyang9/BPR-pytorch}}~\cite{rendle2009bpr} is a collaborative filtering method based on matrix factorization with BPR ranking loss.
10) \textbf{LightGCN}\footnote{\url{https://github.com/gusye1234/LightGCN-PyTorch}}~\cite{he2020lightgcn} leverages high-order neighbors information to enhance the user and item representations.
11) \textbf{SGL}\footnote{\url{https://github.com/wujcan/SGL-Torch}}~\cite{wu2021sgl} is a self-supervised learning method, which conducts graph data augmentation for robust representation learning.

\textbf{Evaluation Metrics.} We adopt the full-ranking protocol to evaluate the top-K recommendation performance using two widely used metrics: Recall@K and NDCG@K with $K=\{10, 20\}$.

\begin{figure*}[t]
\setlength{\abovecaptionskip}{-0.20cm}
\setlength{\belowcaptionskip}{-0.4cm}
  \centering 
  \hspace{-0.7in}
  \subfigure{
    \includegraphics[width=1.62in]{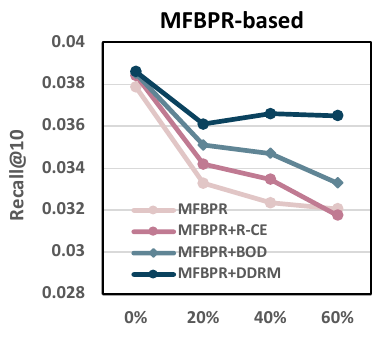}} 
  \hspace{-0.105in}
  \subfigure{
     \includegraphics[width=1.62in]{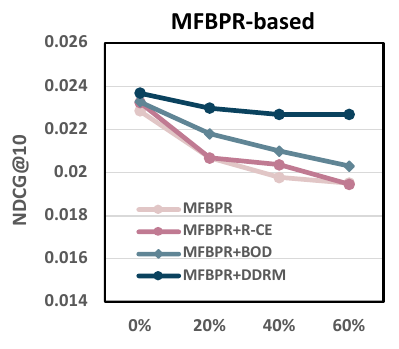}} 
\hspace{-0.105in}
  \subfigure{
    \includegraphics[width=1.7in]{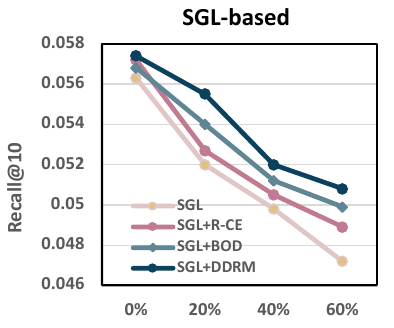}} 
\hspace{-0.105in}
  \subfigure{
    \includegraphics[width=1.67in]{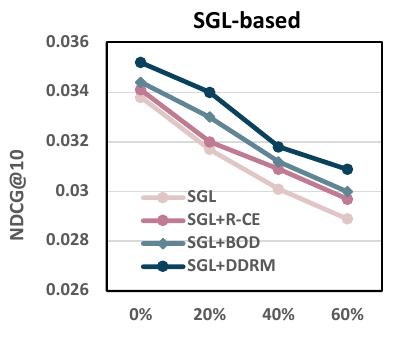}} 
  \hspace{-0.7in} 
\caption{Performance comparison of noisy training with random noises in Yelp.}
  \label{fig:random_noise}
  \vspace{-0.20cm}
\end{figure*}

\subsubsection{\textbf{Hyper-parameter Settings.}} 
We fix the embedding size at 64 to maintain fairness when evaluating different methods.
For model-agnostic methods, we initially determine the optimal hyper-parameters of the three backend models according to their default settings.  
Subsequently, we maintain the backend models' hyper-parameters at their optimal and adjust only the specific denoising parameters, as per the original papers. For non-model-agnostic baselines, hyper-parameters are tuned within their default ranges.

Regarding our proposed DDRM, we have six hyper-parameters in total: the diffusion steps $T$, the noise lower bound $\alpha_{\min}$, the noise upper bound $\alpha_{\max}$, the noise scale $s$ and loss balance factor $\lambda$ and denoising weight factor $\sigma$. In detail, $T$ is tuned within $T=\{10, 20, \dots, 60\}$. As for the noise-related parameters $\alpha_{\min}$, $\alpha_{\max}$ and $s$, we explore the combinations in $\{1e-4, 1e-3\}$, $\{1e-3, 1e-2\}$ and $\{1e-4, 1e-3\}$, respectively. For the loss-related parameters, loss balance factor $\lambda$ and denoising weight factor $\gamma$ are tuned within $\{0.1,0.2,\dots,0.6\}$ and $\{0, 0.05, 0.1,0.2,\dots,0.9\}$, respectively.

\subsection{Overall Performance (RQ1)}

We conduct comprehensive experiments in natural noise setting to compare DDRM's performance with other referenced baselines. The results, illustrated in Table~\ref{tab:main_exp}, yield several key observations:
\begin{itemize}[leftmargin=*]
    \item DDRM mostly outperforms backend models and other model-agnostic denoising methods across all three datasets. This superior performance can be attributed to DDRM's denoising diffusion process, which enhances robust representation learning through multi-step denoising.
    \item The performance of DeCA is not consistently better than the backend model. Two potential reasons emerge: 1) DeCA operates under the presumption that distinct models yield analogous predictions on clean data but deviate on noisy data. This assumption may not consistently hold true across our datasets. 2) DeCA's training process involves four models optimized concurrently, which potentially induces instability.
    \item DiffRec consistently exhibits commendable performance across all three datasets, thereby highlighting the adeptness of diffusion models in denoising. DDRM, more flexible than DiffRec owing to its model-agnostic identity, can be deployed on any recommender model with user and item embeddings. What's more, as for DiffRec, it requires to perform prediction tasks on all candidate items for a given user, resulting in high computational costs. DDRM only necessitates generating one single ideal item at the embedding level, and calculating the score between generated item embeddings and candidate item embeddings, which is more efficient. Furthermore, DDRM bolsters suboptimal models to perform comparably with, or even surpass DiffRec, thereby denoting its tangible enhancements upon the backend model.
\end{itemize}

\subsection{In-depth Analysis}

\begin{figure}[t]
\setlength{\abovecaptionskip}{-0.20cm}
\setlength{\belowcaptionskip}{-0.7cm}
  \centering 
  \hspace{-0.2in}
  \subfigure{
    \includegraphics[width=1.4in]{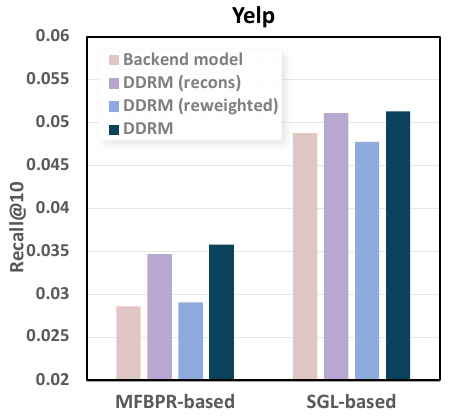}} 
  \hspace{-0.02in}
  \subfigure{
     \includegraphics[width=1.4in]{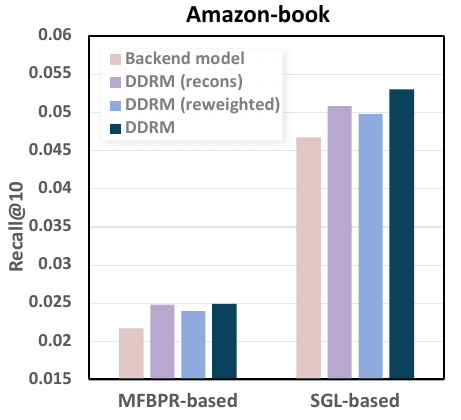}} 
\caption{Contributions of reconstruction loss and reweighted loss to DDRM compared with backend models.}
  \label{fig:abl}
\end{figure}

\subsubsection{\textbf{Random Noisy Training (RQ1).}} 
We conduct random noisy training to evaluate the noise resistance capability of DDRM, comparing it with the two most competitive model-agnostic methods, R-CE and BOD, along with the backend model. 
The proportion of noise in our training settings spanned from 0\% to 60\%.
We report the results in Figure~\ref{fig:random_noise}. 
Similar results are seen with Amazon-book and ML-1M, but figures are omitted for brevity.
The results show that:
1) As the noise ratio increases, there is an overall declining trend in the performance of the backend model, R-CE, BOD, and DDRM.
This decline is attributed to the intensifying corruption of data due to the escalating noise level, making it challenging to discern genuine user preferences.
2) DDRM consistently outperforms both the backend model and R-CE in different noise ratio settings. This emphasizes DDRM's commendable noise resistance, attributed to its robust representation learned through the diffusion process. 

\begin{figure}
\setlength{\abovecaptionskip}{-0.20cm}
\setlength{\belowcaptionskip}{-0.2cm}
  \centering 
  \hspace{-0.2in}
  \subfigure{
    \includegraphics[width=1.4in]{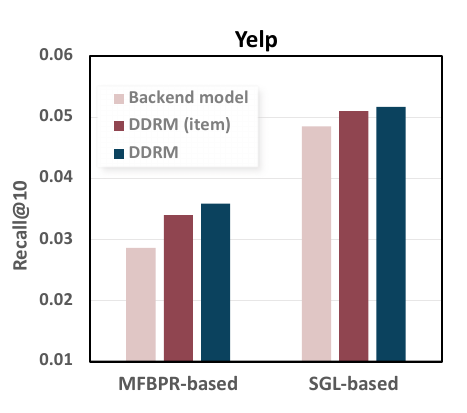}} 
  \hspace{-0.02in}
  \subfigure{
     \includegraphics[width=1.4in]{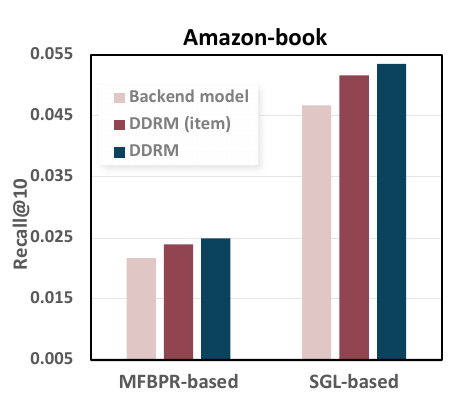}} 
\caption{Contributions of user and item denoising modules to DDRM compared with backend models.}
  \label{fig:abl_module}
\end{figure}

\begin{figure*}[t]
\setlength{\abovecaptionskip}{-0.10cm}
\setlength{\belowcaptionskip}{-0.3cm}
  \centering 
  \hspace{-0.7in}
  \subfigure{
    \includegraphics[width=1.67in]{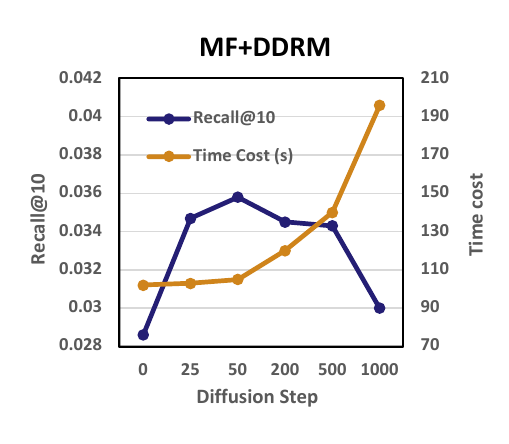}} 
  \hspace{-0.105in}
  \subfigure{
     \includegraphics[width=1.67in]{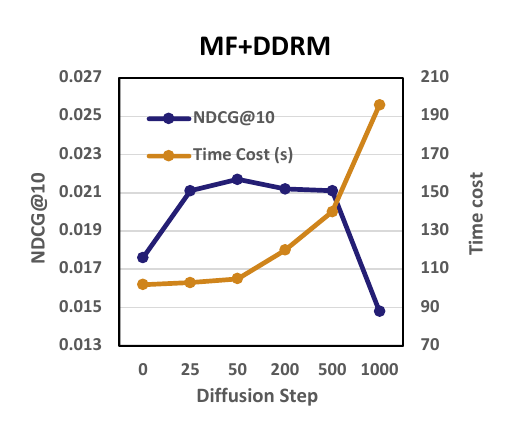}} 
\hspace{-0.105in}
  \subfigure{
    \includegraphics[width=1.67in]{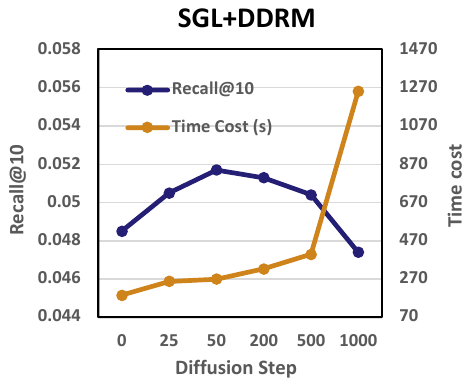}} 
\hspace{-0.105in}
  \subfigure{
    \includegraphics[width=1.67in]{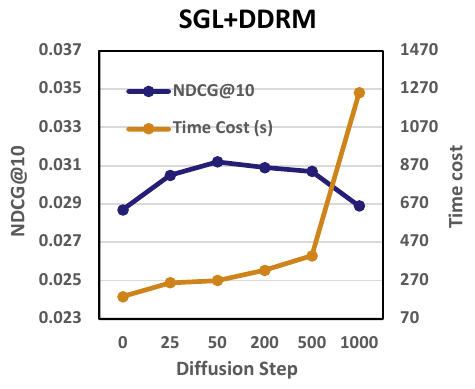}} 
  \hspace{-0.7in} 
\caption{Performance and inference time cost comparison of using different diffusion steps in the DDRM in Yelp.}
  \label{fig:infer_time}
  \vspace{-0.20cm}
\end{figure*}

\subsubsection{\textbf{Ablation Study (RQ2).}}
\label{sec:ablation}
We execute ablation study to analyze DDRM from two distinct angles: loss perspective and module perspective.

\vspace{3pt}
\noindent{\textbf{$\bullet$ Loss Perspective.}} 
We assess the distinct contributions of the reconstruction loss from the reverse process and the reweighted loss from the extension, with outcomes depicted in Figure~\ref{fig:abl} for Yelp and Amazon-book (omitting ML-1M due to similar trends). 
We select the classic model MF and the graph model SGL which has uniformly commendable performance as our backend models.
Based on these results, we can find that:
1) DDRM with reconstruction loss consistently outperforms the backend model, underscoring the efficacy of embedding denoising in DDRM.
2) DDRM with reweighted loss demonstrates overall effectiveness, yet its performance exhibits variability across different datasets and backend models. 
Notably, there exist instances where it falls short of the backend model’s performance.
Furthermore, while improvements are observed in most cases, they are generally less substantial compared to those achieved by the DDRM with reconstruction loss. 
This suggests that the underlying assumptions of data cleaning methods (\cf Sec~\ref{sec:introduction}) may not always hold true.
3) Overall, while reconstruction and reweighted losses jointly contribute to DDRM’s effectiveness, the latter's contributions are notably milder than those of the former.

\vspace{3pt}
\noindent{\textbf{$\bullet$ Module Perspective.}} 
To explore the impact of user denoising module and item denoising module, we conduct additional experiments on the Yelp and Amazon-book dataset only keeping the item denoising module of DDRM, with performance shown in Figure~\ref{fig:abl_module}. 
It is important to note that experiments focusing solely on user denoising were not feasible in our work since we need to denoise item embeddings to generate ideal items for users in the inference phase.
The results indicate a performance decrease compared to the full DDRM approach, suggesting that noise is present in user embeddings and necessitates user denoising. However, the performance with only item denoising still surpasses that of the backend models alone, which implies that the item denoising module contributes effectively to noise reduction. Therefore, it can be inferred that noise exists in both user and item representations, which prove the effectiveness and necessity of both user denoising module and item denoising module.

\begin{table}[]
\setlength{\abovecaptionskip}{0cm}
\setlength{\belowcaptionskip}{0.2cm}
\caption{Performance of different design variations in Yelp. Bold signifies the best performance among listed methods.}
\setlength{\tabcolsep}{3mm}{
\resizebox{\linewidth}{!}{
\label{tab:backbone}
\begin{tabular}{c|cccc}
\hline
\textbf{Model}& \textbf{R@10} & \textbf{R@20} & \textbf{N@10} & \textbf{N@20} \\ \hline
\textbf{SGL}                & 0.0488        & 0.0841        & 0.0290         & 0.0397        \\
\textbf{DDRM}               & \textbf{0.0517}        & \textbf{0.0860}         & \textbf{0.0312}        & \textbf{0.0415}        \\
\textbf{DDRM (Transformer)} & 0.0469        & 0.0786        & 0.0282        & 0.0378        \\
\textbf{DDRM (Noise Inference)}& 0.0213        & 0.0364        & 0.0133        & 0.0179        \\
 \textbf{DDRM (Schedule: Linear)}& 0.0496& 0.0834& 0.0303&0.0404\\
 \textbf{DDRM (Schedule: Cosine)}& 0.0502& 0.0834& 0.0302&0.0406\\
 \textbf{DDRM (Schedule: Binomial)}& 0.0491& 0.0828& 0.0295&0.0397\\
 \hline
\end{tabular}
}}
\end{table}

\subsubsection{\textbf{Design Variations (RQ3).}} 
We explore various design variations to validate the performance of our proposed DDRM. The performance of different designs is shown in Table~\ref{tab:backbone}, noting that the default backend model is SGL since it has uniformly competitive performance across three datasets. The detailed analyses of each backbone design are provided below.

\vspace{3pt}
\noindent{\textbf{$\bullet$ Architecture.}} 
In this variation, we substitute the reconstruction MLP within the denoising module with the standard Transformer architecture. Our objective is to investigate whether the inclusion of a more intricate attention mechanism could bolster performance.
Specifically, the collaborative information and step embedding are designated as the query and key in the Transformer, respectively, while the embedding requiring denoising is assigned as the value~\cite{vaswani2017attention}. 
However, the results reveales the Transformer's inferiority to the MLP in this context.  
A likely rationale for this outcome lies in the inductive bias introduced by the additional model structures in the Transformer, which is inappropriate in the denoising context, thus rendering suboptimal performance in reconstructing these embeddings. In contrast, the simplicity of the MLP proves to be highly efficient for the task at hand.

\vspace{3pt}
\noindent{\textbf{$\bullet$ Noise Inference.}}
As inspired by~\cite{li2023diffurec}, another approach we explored is denoising the embedding exclusively from pure noise instead of noisy average embeddings $\bm{\bar{e}_T^i}$ (\cf Section~\ref{sec:inference}) during the inference phase.
This method highly depends on diffusion's impressive generation capabilities. Nonetheless, this leads to a substantial decline in performance, corroborating the assertions made in Section~\ref{sec:introduction}. 
Using pure noise as the starting point will increase the difficulty of the denoising process, primarily due to the absence of personalization which is crucial for regulating the generation of the ideal item in the recommendation scenario.



\vspace{3pt}
\noindent{\textbf{$\bullet$ Noise Schedule.}}
We also examine the effect of using various noise schedules in the forward process of DDRM. 
Different types of noise schedules control how to add noise iteratively to the embeddings in the forward process, significantly influencing the denoising process. 
While the \textit{linear variance} schedule is chosen for DDRM, we also experimented with three alternatives: the \textit{linear}, \textit{cosine}, and \textit{binomial} schedules~\cite{ho2020denoising}. The findings in Table~\ref{tab:backbone} revealed that the \textit{linear variance} schedule surpasses the others. 
Its superiority lies in facilitating a smoother and more gradual noise injection in the embeddings, which improves the stability in the training phase compared with the \textit{cosine} and \textit{binomial} schedules.
Furthermore, this schedule controls the maximum noise level injected into embeddings, which helps to preserve essential personalized information after the forward process compared with the \textit{linear} schedule.

\begin{figure*}[t]
\setlength{\abovecaptionskip}{-0.10cm}
\setlength{\belowcaptionskip}{-0.3cm}
  \centering 
  \hspace{-0.7in}
  \subfigure{
    \includegraphics[width=1.67in]{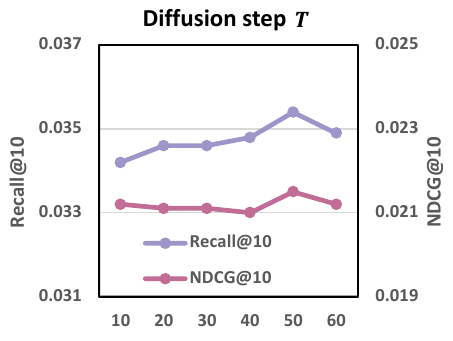}} 
  \hspace{-0.105in}
  \subfigure{
     \includegraphics[width=1.65in]{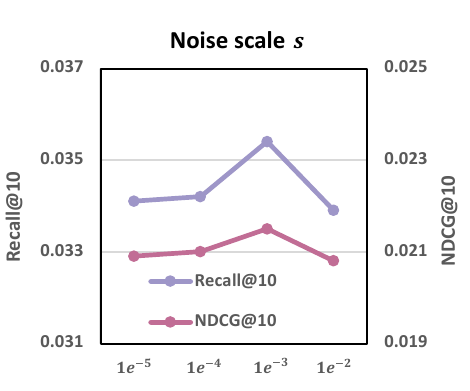}} 
\hspace{-0.105in}
  \subfigure{
    \includegraphics[width=1.67in]{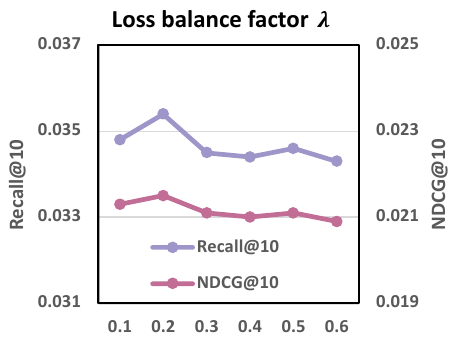}}  
\hspace{-0.105in}
  \subfigure{
    \includegraphics[width=1.67in]{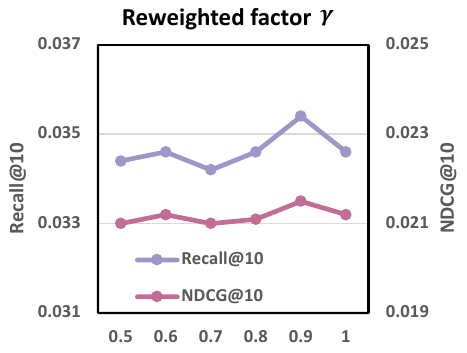}} 
  \hspace{-0.7in} 
\caption{Hyper-parameter analysis in MFBPR-based DDRM for Yelp dataset.}
  \label{fig:hyper}
  \vspace{-0.20cm}
\end{figure*}

\subsubsection{\textbf{Inference Step Analysis (RQ3).}} 
We also delve into the impact of the number of diffusion steps on performance and time efficiency during inference, as depicted in Figure~\ref{fig:infer_time}. 
The findings indicate that optimal performance can be achieved with fewer than 100 diffusion steps. This contrasts with traditional diffusion models commonly employed in generative tasks, which often require up to 1000 steps. 
This is because these traditional models aim to reverse from pure noise during the inference, which need more steps to reverse in order to get high-quality image. DDRM, however, starts from embeddings that already contain personalized information which reflect users’ preference. Therefore, DDRM needs fewer steps to reverse compared with the traditional diffusion models, which saves the computation and time cost.

\subsubsection{\textbf{Hyper-parameters Analysis.}} 
For a more nuanced understanding, we select certain sensitive hyper-parameters, adjusting them within the ranges delineated in Section~\ref{sec:setting}. The outcomes of these experiments are visually represented in Figure~\ref{fig:hyper}. From our observations:
1) With an increase in the diffusion step $T$ and noise scale $s$, DDRM's performance initially rises. This is attributed to an enhanced noise diversity in the data, enabling the model to foster more robust representations. Nonetheless, overly extending the diffusion steps and noise scale adversely affects performance, compromising the model's personalization capabilities. Hence, it becomes imperative to judiciously determine the optimal diffusion step and noise scale to harness peak performance.
2) The loss balance factor $\lambda$ influences DDRM performance by mediating the focus between recommendation and reconstruction tasks. While an increase in $\lambda$ initially bolsters performance by prioritizing embedding denoising, too high a value risks neglecting the core recommendation task, undermining overall performance.
3) Appropriate selection of $\gamma$ is crucial for effective denoising as higher values may filter out clean samples, while lower values may not comprehensively filter noisy samples.

\section{Related Work}
\label{sec:related_work}


\textbf{$\bullet$ Denoising Implicit Feedback.} 
Addressing the noise in implicit feedback, caused by false-positive interactions, has seen approaches primarily from data cleaning perspective~\cite{wang2021denoising,wang2022robust,2011wbpr,wang2023efficient} and model perspective~\cite{wu2016collaborative,jiang2023adaptive}. 
Data cleaning methods often rely on specific assumptions to directly eliminate noisy interactions(\eg WBPR~\cite{2011wbpr}) or assign lower weights to such samples(\eg T-CE~\cite{wang2021denoising}, R-CE~\cite{wang2021denoising}, DeCA~\cite{wang2022robust}), resulting in limited adaptability across different backend models and datasets. 
Model perspective methods, on the other hand, focus on enhancing the noise resistance of recommender model. 
Specifically, some graph-based models (\eg SGL~\cite{wu2021sgl}, AdaGCL~\cite{jiang2023adaptive}) employ data augmentation to enhance data diversity and then regulate models to learn more robust representation. 
Additionally, some auto-encoder based models (\eg CDAE~\cite{wu2016collaborative}) intentionally corrupt users' interactions by introducing different types of noise during training, and then attempt to reconstruct the original clean data using simple auto-encoders. 
However, these model perspective methods expect models to capture the complex noise distribution in the real world, which might pose a considerable challenge for the neural networks. 

\noindent\textbf{$\bullet$ Generative Recommendation.} 
Generative models, notably Generative Adversarial Networks (GANs)~\cite{wang2017irgan,gao2021recommender,jin2020sampling} and Variational Autoencoders (VAEs)~\cite{ma2019learning,liang2018variational,zhang2017autosvd++}, are pivotal for personalized recommendations but face structural limitations~\cite{sohl2015deep, kingma2016improved}. Emergent diffusion models, providing enhanced stability and representation compared with GANs and VAEs, have been recently explored in recommendation contexts~\cite{wu2019neural,WuLSHGW22,ma2024plug,jiang2023diffkg,chen2023adversarial}. 
While models like CODIGEM~\cite{walker2022recommendation} and DiffRec~\cite{diffrec} utilize diffusion models for inferring user preferences by modeling the distribution of users' interaction probabilities, other research~\cite{li2023diffurec,wang2023conditional,du2023sequential,liu2023diffusion} targets content generation at the embedding level, akin to our DDRM. 
For example, DiffuRec~\cite{li2023diffurec} and CDDRec~\cite{wang2023conditional} corrupt target item representations into pure noise in the forward process, subsequently reconstructing them condition on users' historical interaction sequences.
DiffuASR~\cite{liu2023diffusion} uses diffusion models to generate new item sequences, mitigating data sparsity issues.
However, these methods primarily concentrate on sequential recommendation and tend to overlook the presence of natural false-positive interactions in implicit feedback. 
In contrast, our proposed DDRM employs diffusion models to denoise implicit feedback, contributing to more robust representation learning.

\section{Conclusion and Future Work}
\label{sec:conclusion}
In this work, we proposed the Denoising Diffusion Recommender Model (DDRM), a plug-in model to bolster robust representation learning amidst noisy feedback for existing recommender models. 
Given user and item embeddings from any recommender models, DDRM proactively injects Gaussian noises into the embeddings and then iteratively removes noise in the reverse process.
To guide the reverse process, we designed a denoising module to encode collaborative information as guidance. 
During the inference phase, we utilized the average embeddings of users' historically liked items as the starting point to generate an ideal item embedding, and then ground this embedding to existing item candidates as the recommendation for a user.
Extensive experiment results demonstrate the superiority of DDRM compared with other competitive baselines. 

Future enhancements for DDRM may involve: 
1) enriching the denoising module with more well-designed neural networks, and
2) exploring adaptations of DDRM for a broader range of recommendation tasks, such as sequential and bundle recommendations.

\clearpage

{
\tiny
\bibliographystyle{ACM-Reference-Format}
\balance
\bibliography{bibfile}


\begin{thebibliography}{56}


\ifx \showCODEN    \undefined \def \showCODEN     #1{\unskip}     \fi
\ifx \showDOI      \undefined \def \showDOI       #1{#1}\fi
\ifx \showISBNx    \undefined \def \showISBNx     #1{\unskip}     \fi
\ifx \showISBNxiii \undefined \def \showISBNxiii  #1{\unskip}     \fi
\ifx \showISSN     \undefined \def \showISSN      #1{\unskip}     \fi
\ifx \showLCCN     \undefined \def \showLCCN      #1{\unskip}     \fi
\ifx \shownote     \undefined \def \shownote      #1{#1}          \fi
\ifx \showarticletitle \undefined \def \showarticletitle #1{#1}   \fi
\ifx \showURL      \undefined \def \showURL       {\relax}        \fi
\providecommand\bibfield[2]{#2}
\providecommand\bibinfo[2]{#2}
\providecommand\natexlab[1]{#1}
\providecommand\showeprint[2][]{arXiv:#2}

\bibitem[Chen et~al\mbox{.}(2022)]%
        {chen2022denoising}
\bibfield{author}{\bibinfo{person}{Huiyuan Chen}, \bibinfo{person}{Yusan Lin}, \bibinfo{person}{Menghai Pan}, \bibinfo{person}{Lan Wang}, \bibinfo{person}{Chin-Chia~Michael Yeh}, \bibinfo{person}{Xiaoting Li}, \bibinfo{person}{Yan Zheng}, \bibinfo{person}{Fei Wang}, {and} \bibinfo{person}{Hao Yang}.} \bibinfo{year}{2022}\natexlab{}.
\newblock \showarticletitle{Denoising self-attentive sequential recommendation}. In \bibinfo{booktitle}{\emph{RecSys}}. \bibinfo{publisher}{ACM}, \bibinfo{pages}{92--101}.
\newblock


\bibitem[Chen et~al\mbox{.}(2021)]%
        {chen2021autodebias}
\bibfield{author}{\bibinfo{person}{Jiawei Chen}, \bibinfo{person}{Hande Dong}, \bibinfo{person}{Yang Qiu}, \bibinfo{person}{Xiangnan He}, \bibinfo{person}{Xin Xin}, \bibinfo{person}{Liang Chen}, \bibinfo{person}{Guli Lin}, {and} \bibinfo{person}{Keping Yang}.} \bibinfo{year}{2021}\natexlab{}.
\newblock \showarticletitle{AutoDebias: Learning to debias for recommendation}. In \bibinfo{booktitle}{\emph{SIGIR}}. \bibinfo{publisher}{ACM}, \bibinfo{pages}{21--30}.
\newblock


\bibitem[Chen et~al\mbox{.}(2023)]%
        {chen2023adversarial}
\bibfield{author}{\bibinfo{person}{Lijian Chen}, \bibinfo{person}{Wei Yuan}, \bibinfo{person}{Tong Chen}, \bibinfo{person}{Quoc Viet~Hung Nguyen}, \bibinfo{person}{Lizhen Cui}, {and} \bibinfo{person}{Hongzhi Yin}.} \bibinfo{year}{2023}\natexlab{}.
\newblock \showarticletitle{Adversarial Item Promotion on Visually-Aware Recommender Systems by Guided Diffusion}.
\newblock \bibinfo{journal}{\emph{arXiv:2312.15826}} (\bibinfo{year}{2023}).
\newblock


\bibitem[Croitoru et~al\mbox{.}(2022)]%
        {croitoru2022diffusion}
\bibfield{author}{\bibinfo{person}{Florinel-Alin Croitoru}, \bibinfo{person}{Vlad Hondru}, \bibinfo{person}{Radu~Tudor Ionescu}, {and} \bibinfo{person}{Mubarak Shah}.} \bibinfo{year}{2022}\natexlab{}.
\newblock \showarticletitle{Diffusion models in vision: A survey}.
\newblock \bibinfo{journal}{\emph{arXiv:2209.04747}} (\bibinfo{year}{2022}).
\newblock


\bibitem[Ding et~al\mbox{.}(2018)]%
        {ding2018improved}
\bibfield{author}{\bibinfo{person}{Jingtao Ding}, \bibinfo{person}{Fuli Feng}, \bibinfo{person}{Xiangnan He}, \bibinfo{person}{Guanghui Yu}, \bibinfo{person}{Yong Li}, {and} \bibinfo{person}{Depeng Jin}.} \bibinfo{year}{2018}\natexlab{}.
\newblock \showarticletitle{An improved sampler for bayesian personalized ranking by leveraging view data}. In \bibinfo{booktitle}{\emph{WWW}}. \bibinfo{publisher}{ACM}, \bibinfo{pages}{13--14}.
\newblock


\bibitem[Ding et~al\mbox{.}(2019)]%
        {ding2019sampler}
\bibfield{author}{\bibinfo{person}{Jingtao Ding}, \bibinfo{person}{Guanghui Yu}, \bibinfo{person}{Xiangnan He}, \bibinfo{person}{Fuli Feng}, \bibinfo{person}{Yong Li}, {and} \bibinfo{person}{Depeng Jin}.} \bibinfo{year}{2019}\natexlab{}.
\newblock \showarticletitle{Sampler design for bayesian personalized ranking by leveraging view data}.
\newblock \bibinfo{journal}{\emph{TKDE}} (\bibinfo{year}{2019}), \bibinfo{pages}{667--681}.
\newblock


\bibitem[Du et~al\mbox{.}(2023)]%
        {du2023sequential}
\bibfield{author}{\bibinfo{person}{Hanwen Du}, \bibinfo{person}{Huanhuan Yuan}, \bibinfo{person}{Zhen Huang}, \bibinfo{person}{Pengpeng Zhao}, {and} \bibinfo{person}{Xiaofang Zhou}.} \bibinfo{year}{2023}\natexlab{}.
\newblock \showarticletitle{Sequential Recommendation with Diffusion Models}.
\newblock \bibinfo{journal}{\emph{arXiv:2304.04541}}.
\newblock


\bibitem[Fan et~al\mbox{.}(2023)]%
        {fan2023graph}
\bibfield{author}{\bibinfo{person}{Ziwei Fan}, \bibinfo{person}{Ke Xu}, \bibinfo{person}{Zhang Dong}, \bibinfo{person}{Hao Peng}, \bibinfo{person}{Jiawei Zhang}, {and} \bibinfo{person}{Philip~S Yu}.} \bibinfo{year}{2023}\natexlab{}.
\newblock \showarticletitle{Graph Collaborative Signals Denoising and Augmentation for Recommendation}. In \bibinfo{booktitle}{\emph{SIGIR}}. \bibinfo{publisher}{ACM}, \bibinfo{pages}{2037--2041}.
\newblock


\bibitem[Gantner et~al\mbox{.}(2011)]%
        {2011wbpr}
\bibfield{author}{\bibinfo{person}{Zeno Gantner}, \bibinfo{person}{Lucas Drumond}, \bibinfo{person}{Christoph Freudenthaler}, {and} \bibinfo{person}{Lars Schmidt-Thieme}.} \bibinfo{year}{2011}\natexlab{}.
\newblock \showarticletitle{Personalized Ranking for Non-Uniformly Sampled Items}. In \bibinfo{booktitle}{\emph{KDDCUP}}. \bibinfo{publisher}{JMLR}, \bibinfo{pages}{231–247}.
\newblock


\bibitem[Gao et~al\mbox{.}(2021)]%
        {gao2021recommender}
\bibfield{author}{\bibinfo{person}{Min Gao}, \bibinfo{person}{Junwei Zhang}, \bibinfo{person}{Junliang Yu}, \bibinfo{person}{Jundong Li}, \bibinfo{person}{Junhao Wen}, {and} \bibinfo{person}{Qingyu Xiong}.} \bibinfo{year}{2021}\natexlab{}.
\newblock \showarticletitle{Recommender systems based on generative adversarial networks: A problem-driven perspective}.
\newblock \bibinfo{journal}{\emph{Inf. Sci.}} (\bibinfo{year}{2021}), \bibinfo{pages}{1166--1185}.
\newblock


\bibitem[Gao et~al\mbox{.}(2022)]%
        {gao2022self}
\bibfield{author}{\bibinfo{person}{Yunjun Gao}, \bibinfo{person}{Yuntao Du}, \bibinfo{person}{Yujia Hu}, \bibinfo{person}{Lu Chen}, \bibinfo{person}{Xinjun Zhu}, \bibinfo{person}{Ziquan Fang}, {and} \bibinfo{person}{Baihua Zheng}.} \bibinfo{year}{2022}\natexlab{}.
\newblock \showarticletitle{Self-guided learning to denoise for robust recommendation}. In \bibinfo{booktitle}{\emph{SIGIR}}. \bibinfo{publisher}{ACM}, \bibinfo{pages}{1412--1422}.
\newblock


\bibitem[Guo et~al\mbox{.}(2023)]%
        {guo2023towards}
\bibfield{author}{\bibinfo{person}{Shuyu Guo}, \bibinfo{person}{Shuo Zhang}, \bibinfo{person}{Weiwei Sun}, \bibinfo{person}{Pengjie Ren}, \bibinfo{person}{Zhumin Chen}, {and} \bibinfo{person}{Zhaochun Ren}.} \bibinfo{year}{2023}\natexlab{}.
\newblock \showarticletitle{Towards explainable conversational recommender systems}. In \bibinfo{booktitle}{\emph{SIGIR}}. \bibinfo{publisher}{ACM}, \bibinfo{pages}{2786--2795}.
\newblock


\bibitem[He et~al\mbox{.}(2020)]%
        {he2020lightgcn}
\bibfield{author}{\bibinfo{person}{Xiangnan He}, \bibinfo{person}{Kuan Deng}, \bibinfo{person}{Xiang Wang}, \bibinfo{person}{Yan Li}, \bibinfo{person}{Yongdong Zhang}, {and} \bibinfo{person}{Meng Wang}.} \bibinfo{year}{2020}\natexlab{}.
\newblock \showarticletitle{Lightgcn: Simplifying and powering graph convolution network for recommendation}. In \bibinfo{booktitle}{\emph{SIGIR}}. \bibinfo{pages}{639--648}.
\newblock


\bibitem[Ho et~al\mbox{.}(2020)]%
        {ho2020denoising}
\bibfield{author}{\bibinfo{person}{Jonathan Ho}, \bibinfo{person}{Ajay Jain}, {and} \bibinfo{person}{Pieter Abbeel}.} \bibinfo{year}{2020}\natexlab{}.
\newblock \showarticletitle{Denoising diffusion probabilistic models}. In \bibinfo{booktitle}{\emph{NeurIPS}}. \bibinfo{publisher}{Curran Associates, Inc.}, \bibinfo{pages}{6840--6851}.
\newblock


\bibitem[Hoogeboom et~al\mbox{.}(2021)]%
        {hoogeboom2021argmax}
\bibfield{author}{\bibinfo{person}{Emiel Hoogeboom}, \bibinfo{person}{Didrik Nielsen}, \bibinfo{person}{Priyank Jaini}, \bibinfo{person}{Patrick Forr{\'e}}, {and} \bibinfo{person}{Max Welling}.} \bibinfo{year}{2021}\natexlab{}.
\newblock \showarticletitle{Argmax flows and multinomial diffusion: Learning categorical distributions}. \bibinfo{publisher}{Curran Associates, Inc.}, \bibinfo{pages}{12454--12465}.
\newblock


\bibitem[Huang et~al\mbox{.}(2023)]%
        {huang2023mdm}
\bibfield{author}{\bibinfo{person}{Lei Huang}, \bibinfo{person}{Hengtong Zhang}, \bibinfo{person}{Tingyang Xu}, {and} \bibinfo{person}{Ka-Chun Wong}.} \bibinfo{year}{2023}\natexlab{}.
\newblock \showarticletitle{Mdm: Molecular diffusion model for 3d molecule generation}. In \bibinfo{booktitle}{\emph{AAAI}}. \bibinfo{publisher}{AAAI press}, \bibinfo{pages}{5105--5112}.
\newblock


\bibitem[Jiang et~al\mbox{.}(2023a)]%
        {jiang2023adaptive}
\bibfield{author}{\bibinfo{person}{Yangqin Jiang}, \bibinfo{person}{Chao Huang}, {and} \bibinfo{person}{Lianghao Huang}.} \bibinfo{year}{2023}\natexlab{a}.
\newblock \showarticletitle{Adaptive Graph Contrastive Learning for Recommendation}. In \bibinfo{booktitle}{\emph{KDD}}. \bibinfo{publisher}{ACM}, \bibinfo{pages}{4252–4261}.
\newblock


\bibitem[Jiang et~al\mbox{.}(2023b)]%
        {jiang2023diffkg}
\bibfield{author}{\bibinfo{person}{Yangqin Jiang}, \bibinfo{person}{Yuhao Yang}, \bibinfo{person}{Lianghao Xia}, {and} \bibinfo{person}{Chao Huang}.} \bibinfo{year}{2023}\natexlab{b}.
\newblock \showarticletitle{DiffKG: Knowledge Graph Diffusion Model for Recommendation}.
\newblock \bibinfo{journal}{\emph{arXiv:2312.16890}} (\bibinfo{year}{2023}).
\newblock


\bibitem[Jin et~al\mbox{.}(2020)]%
        {jin2020sampling}
\bibfield{author}{\bibinfo{person}{Binbin Jin}, \bibinfo{person}{Defu Lian}, \bibinfo{person}{Zheng Liu}, \bibinfo{person}{Qi Liu}, \bibinfo{person}{Jianhui Ma}, \bibinfo{person}{Xing Xie}, {and} \bibinfo{person}{Enhong Chen}.} \bibinfo{year}{2020}\natexlab{}.
\newblock \showarticletitle{Sampling-decomposable generative adversarial recommender}. In \bibinfo{booktitle}{\emph{NeurIPS}}. \bibinfo{publisher}{Curran Associates, Inc.}, \bibinfo{pages}{22629--22639}.
\newblock


\bibitem[Kingma et~al\mbox{.}(2016)]%
        {kingma2016improved}
\bibfield{author}{\bibinfo{person}{Durk~P Kingma}, \bibinfo{person}{Tim Salimans}, \bibinfo{person}{Rafal Jozefowicz}, \bibinfo{person}{Xi Chen}, \bibinfo{person}{Ilya Sutskever}, {and} \bibinfo{person}{Max Welling}.} \bibinfo{year}{2016}\natexlab{}.
\newblock \showarticletitle{Improved variational inference with inverse autoregressive flow}. In \bibinfo{booktitle}{\emph{NeurIPS}}. \bibinfo{publisher}{Curran Associates, Inc.}, \bibinfo{pages}{4743–4751}.
\newblock


\bibitem[Li et~al\mbox{.}(2022)]%
        {li2022diffusion}
\bibfield{author}{\bibinfo{person}{Xiang~Lisa Li}, \bibinfo{person}{John Thickstun}, \bibinfo{person}{Ishaan Gulrajani}, \bibinfo{person}{Percy Liang}, {and} \bibinfo{person}{Tatsunori~B Hashimoto}.} \bibinfo{year}{2022}\natexlab{}.
\newblock \showarticletitle{Diffusion-lm improves controllable text generation}.
\newblock \bibinfo{journal}{\emph{arXiv:2205.14217}}.
\newblock


\bibitem[Li et~al\mbox{.}(2023)]%
        {li2023diffurec}
\bibfield{author}{\bibinfo{person}{Zihao Li}, \bibinfo{person}{Aixin Sun}, {and} \bibinfo{person}{Chenliang Li}.} \bibinfo{year}{2023}\natexlab{}.
\newblock \showarticletitle{DiffuRec: A Diffusion Model for Sequential Recommendation}.
\newblock \bibinfo{journal}{\emph{arXiv:2304.00686}}.
\newblock


\bibitem[Liang et~al\mbox{.}(2018)]%
        {liang2018variational}
\bibfield{author}{\bibinfo{person}{Dawen Liang}, \bibinfo{person}{Rahul~G Krishnan}, \bibinfo{person}{Matthew~D Hoffman}, {and} \bibinfo{person}{Tony Jebara}.} \bibinfo{year}{2018}\natexlab{}.
\newblock \showarticletitle{Variational Autoencoders for Collaborative Filtering}. In \bibinfo{booktitle}{\emph{WWW}}. \bibinfo{publisher}{ACM}, \bibinfo{pages}{689--698}.
\newblock


\bibitem[Lin et~al\mbox{.}(2023)]%
        {lin2023autodenoise}
\bibfield{author}{\bibinfo{person}{Weilin Lin}, \bibinfo{person}{Xiangyu Zhao}, \bibinfo{person}{Yejing Wang}, \bibinfo{person}{Yuanshao Zhu}, {and} \bibinfo{person}{Wanyu Wang}.} \bibinfo{year}{2023}\natexlab{}.
\newblock \showarticletitle{AutoDenoise: Automatic Data Instance Denoising for Recommendations}. In \bibinfo{booktitle}{\emph{WWW}}. \bibinfo{publisher}{ACM}, \bibinfo{pages}{1003--1011}.
\newblock


\bibitem[Liu et~al\mbox{.}(2023)]%
        {liu2023diffusion}
\bibfield{author}{\bibinfo{person}{Qidong Liu}, \bibinfo{person}{Fan Yan}, \bibinfo{person}{Xiangyu Zhao}, \bibinfo{person}{Zhaocheng Du}, \bibinfo{person}{Huifeng Guo}, \bibinfo{person}{Ruiming Tang}, {and} \bibinfo{person}{Feng Tian}.} \bibinfo{year}{2023}\natexlab{}.
\newblock \showarticletitle{Diffusion Augmentation for Sequential Recommendation}.
\newblock \bibinfo{journal}{\emph{arXiv:2309.12858}}.
\newblock


\bibitem[Luo(2022)]%
        {luo2022understanding}
\bibfield{author}{\bibinfo{person}{Calvin Luo}.} \bibinfo{year}{2022}\natexlab{}.
\newblock \showarticletitle{Understanding diffusion models: A unified perspective}.
\newblock \bibinfo{journal}{\emph{arXiv:2208.11970}}.
\newblock


\bibitem[Ma et~al\mbox{.}(2024)]%
        {ma2024plug}
\bibfield{author}{\bibinfo{person}{Haokai Ma}, \bibinfo{person}{Ruobing Xie}, \bibinfo{person}{Lei Meng}, \bibinfo{person}{Xin Chen}, \bibinfo{person}{Xu Zhang}, \bibinfo{person}{Leyu Lin}, {and} \bibinfo{person}{Zhanhui Kang}.} \bibinfo{year}{2024}\natexlab{}.
\newblock \showarticletitle{Plug-in Diffusion Model for Sequential Recommendation}.
\newblock \bibinfo{journal}{\emph{arXiv:2401.02913}} (\bibinfo{year}{2024}).
\newblock


\bibitem[Ma et~al\mbox{.}(2019)]%
        {ma2019learning}
\bibfield{author}{\bibinfo{person}{Jianxin Ma}, \bibinfo{person}{Chang Zhou}, \bibinfo{person}{Peng Cui}, \bibinfo{person}{Hongxia Yang}, {and} \bibinfo{person}{Wenwu Zhu}.} \bibinfo{year}{2019}\natexlab{}.
\newblock \showarticletitle{Learning Disentangled Representations for Recommendation}. In \bibinfo{booktitle}{\emph{NeurIPS}}. \bibinfo{publisher}{Curran Associates, Inc.}, \bibinfo{pages}{5712--5723}.
\newblock


\bibitem[Quan et~al\mbox{.}(2023)]%
        {quan2023robust}
\bibfield{author}{\bibinfo{person}{Yuhan Quan}, \bibinfo{person}{Jingtao Ding}, \bibinfo{person}{Chen Gao}, \bibinfo{person}{Lingling Yi}, \bibinfo{person}{Depeng Jin}, {and} \bibinfo{person}{Yong Li}.} \bibinfo{year}{2023}\natexlab{}.
\newblock \showarticletitle{Robust Preference-Guided Denoising for Graph based Social Recommendation}. In \bibinfo{booktitle}{\emph{WWW}}. \bibinfo{publisher}{ACM}, \bibinfo{pages}{1097--1108}.
\newblock


\bibitem[Rendle et~al\mbox{.}(2009)]%
        {rendle2009bpr}
\bibfield{author}{\bibinfo{person}{Steffen Rendle}, \bibinfo{person}{Christoph Freudenthaler}, \bibinfo{person}{Zeno Gantner}, {and} \bibinfo{person}{Lars Schmidt-Thieme}.} \bibinfo{year}{2009}\natexlab{}.
\newblock \showarticletitle{BPR: Bayesian personalized ranking from implicit feedback}. In \bibinfo{booktitle}{\emph{UAI}}. AUAI Press, \bibinfo{pages}{452--461}.
\newblock


\bibitem[Ruiz et~al\mbox{.}(2023)]%
        {ruiz2023dreambooth}
\bibfield{author}{\bibinfo{person}{Nataniel Ruiz}, \bibinfo{person}{Yuanzhen Li}, \bibinfo{person}{Varun Jampani}, \bibinfo{person}{Yael Pritch}, \bibinfo{person}{Michael Rubinstein}, {and} \bibinfo{person}{Kfir Aberman}.} \bibinfo{year}{2023}\natexlab{}.
\newblock \showarticletitle{Dreambooth: Fine tuning text-to-image diffusion models for subject-driven generation}. In \bibinfo{booktitle}{\emph{CVPR}}. \bibinfo{publisher}{IEEE}, \bibinfo{pages}{22500--22510}.
\newblock


\bibitem[Sohl-Dickstein et~al\mbox{.}(2015)]%
        {sohl2015deep}
\bibfield{author}{\bibinfo{person}{Jascha Sohl-Dickstein}, \bibinfo{person}{Eric Weiss}, \bibinfo{person}{Niru Maheswaranathan}, {and} \bibinfo{person}{Surya Ganguli}.} \bibinfo{year}{2015}\natexlab{}.
\newblock \showarticletitle{Deep unsupervised learning using nonequilibrium thermodynamics}. In \bibinfo{booktitle}{\emph{ICML}}. PMLR, \bibinfo{pages}{2256--2265}.
\newblock


\bibitem[Tian et~al\mbox{.}(2022)]%
        {tian2022learning}
\bibfield{author}{\bibinfo{person}{Changxin Tian}, \bibinfo{person}{Yuexiang Xie}, \bibinfo{person}{Yaliang Li}, \bibinfo{person}{Nan Yang}, {and} \bibinfo{person}{Wayne~Xin Zhao}.} \bibinfo{year}{2022}\natexlab{}.
\newblock \showarticletitle{Learning to denoise unreliable interactions for graph collaborative filtering}. In \bibinfo{booktitle}{\emph{SIGIR}}. \bibinfo{publisher}{ACM}, \bibinfo{pages}{122--132}.
\newblock


\bibitem[Vaswani et~al\mbox{.}(2017)]%
        {vaswani2017attention}
\bibfield{author}{\bibinfo{person}{Ashish Vaswani}, \bibinfo{person}{Noam Shazeer}, \bibinfo{person}{Niki Parmar}, \bibinfo{person}{Jakob Uszkoreit}, \bibinfo{person}{Llion Jones}, \bibinfo{person}{Aidan~N Gomez}, \bibinfo{person}{{\L}ukasz Kaiser}, {and} \bibinfo{person}{Illia Polosukhin}.} \bibinfo{year}{2017}\natexlab{}.
\newblock \showarticletitle{Attention is all you need}.
\newblock \bibinfo{journal}{\emph{Advances in neural information processing systems}}  \bibinfo{volume}{30} (\bibinfo{year}{2017}).
\newblock


\bibitem[Vignac et~al\mbox{.}(2022)]%
        {vignac2022digress}
\bibfield{author}{\bibinfo{person}{Clement Vignac}, \bibinfo{person}{Igor Krawczuk}, \bibinfo{person}{Antoine Siraudin}, \bibinfo{person}{Bohan Wang}, \bibinfo{person}{Volkan Cevher}, {and} \bibinfo{person}{Pascal Frossard}.} \bibinfo{year}{2022}\natexlab{}.
\newblock \showarticletitle{Digress: Discrete denoising diffusion for graph generation}. In \bibinfo{booktitle}{\emph{ICLR}}.
\newblock


\bibitem[Walker et~al\mbox{.}(2022)]%
        {walker2022recommendation}
\bibfield{author}{\bibinfo{person}{Joojo Walker}, \bibinfo{person}{Ting Zhong}, \bibinfo{person}{Fengli Zhang}, \bibinfo{person}{Qiang Gao}, {and} \bibinfo{person}{Fan Zhou}.} \bibinfo{year}{2022}\natexlab{}.
\newblock \showarticletitle{Recommendation via Collaborative Diffusion Generative Model}. In \bibinfo{booktitle}{\emph{KSEM}}. \bibinfo{publisher}{Springer}, \bibinfo{pages}{593--605}.
\newblock


\bibitem[Wang et~al\mbox{.}(2017)]%
        {wang2017irgan}
\bibfield{author}{\bibinfo{person}{Jun Wang}, \bibinfo{person}{Lantao Yu}, \bibinfo{person}{Weinan Zhang}, \bibinfo{person}{Yu Gong}, \bibinfo{person}{Yinghui Xu}, \bibinfo{person}{Benyou Wang}, \bibinfo{person}{Peng Zhang}, {and} \bibinfo{person}{Dell Zhang}.} \bibinfo{year}{2017}\natexlab{}.
\newblock \showarticletitle{Irgan: A minimax game for unifying generative and discriminative information retrieval models}. In \bibinfo{booktitle}{\emph{SIGIR}}. \bibinfo{publisher}{ACM}, \bibinfo{pages}{515--524}.
\newblock


\bibitem[Wang et~al\mbox{.}(2021a)]%
        {wang2021denoising}
\bibfield{author}{\bibinfo{person}{Wenjie Wang}, \bibinfo{person}{Fuli Feng}, \bibinfo{person}{Xiangnan He}, \bibinfo{person}{Liqiang Nie}, {and} \bibinfo{person}{Tat-Seng Chua}.} \bibinfo{year}{2021}\natexlab{a}.
\newblock \showarticletitle{Denoising implicit feedback for recommendation}. In \bibinfo{booktitle}{\emph{WSDM}}. \bibinfo{publisher}{ACM}, \bibinfo{pages}{373--381}.
\newblock


\bibitem[Wang et~al\mbox{.}(2023c)]%
        {diffrec}
\bibfield{author}{\bibinfo{person}{Wenjie Wang}, \bibinfo{person}{Yiyan Xu}, \bibinfo{person}{Fuli Feng}, \bibinfo{person}{Xinyu Lin}, \bibinfo{person}{Xiangnan He}, {and} \bibinfo{person}{Tat-Seng Chua}.} \bibinfo{year}{2023}\natexlab{c}.
\newblock \showarticletitle{Diffusion Recommender Model}. In \bibinfo{booktitle}{\emph{SIGIR}}. \bibinfo{publisher}{ACM}, \bibinfo{pages}{832–841}.
\newblock


\bibitem[Wang et~al\mbox{.}(2023b)]%
        {wang2023conditional}
\bibfield{author}{\bibinfo{person}{Yu Wang}, \bibinfo{person}{Zhiwei Liu}, \bibinfo{person}{Liangwei Yang}, {and} \bibinfo{person}{Philip~S. Yu}.} \bibinfo{year}{2023}\natexlab{b}.
\newblock \showarticletitle{Conditional Denoising Diffusion for Sequential Recommendation}.
\newblock \bibinfo{journal}{\emph{arXiv:2304.11433}}.
\newblock


\bibitem[Wang et~al\mbox{.}(2022)]%
        {wang2022robust}
\bibfield{author}{\bibinfo{person}{Yu Wang}, \bibinfo{person}{Xin Xin}, \bibinfo{person}{Zaiqiao Meng}, \bibinfo{person}{Joemon~M Jose}, \bibinfo{person}{Fuli Feng}, {and} \bibinfo{person}{Xiangnan He}.} \bibinfo{year}{2022}\natexlab{}.
\newblock \showarticletitle{Learning Robust Recommenders through Cross-Model Agreement}. In \bibinfo{booktitle}{\emph{WWW}}. \bibinfo{publisher}{ACM}, \bibinfo{pages}{2015–2025}.
\newblock


\bibitem[Wang and Chen(2023)]%
        {wang2023robust}
\bibfield{author}{\bibinfo{person}{Zhenlei Wang} {and} \bibinfo{person}{Xu Chen}.} \bibinfo{year}{2023}\natexlab{}.
\newblock \showarticletitle{Robust Recommendation with Adversarial Gaussian Data Augmentation}. In \bibinfo{booktitle}{\emph{WWW}}. \bibinfo{publisher}{ACM}, \bibinfo{pages}{897--905}.
\newblock


\bibitem[Wang et~al\mbox{.}(2023a)]%
        {wang2023efficient}
\bibfield{author}{\bibinfo{person}{Zongwei Wang}, \bibinfo{person}{Min Gao}, \bibinfo{person}{Wentao Li}, \bibinfo{person}{Junliang Yu}, \bibinfo{person}{Linxin Guo}, {and} \bibinfo{person}{Hongzhi Yin}.} \bibinfo{year}{2023}\natexlab{a}.
\newblock \showarticletitle{Efficient Bi-Level Optimization for Recommendation Denoising}. In \bibinfo{booktitle}{\emph{SIGKDD}}. \bibinfo{publisher}{ACM}, \bibinfo{pages}{2502--2511}.
\newblock


\bibitem[Wang et~al\mbox{.}(2021b)]%
        {wang2021implicit}
\bibfield{author}{\bibinfo{person}{Zitai Wang}, \bibinfo{person}{Qianqian Xu}, \bibinfo{person}{Zhiyong Yang}, \bibinfo{person}{Xiaochun Cao}, {and} \bibinfo{person}{Qingming Huang}.} \bibinfo{year}{2021}\natexlab{b}.
\newblock \showarticletitle{Implicit feedbacks are not always favorable: Iterative relabeled one-class collaborative filtering against noisy interactions}. In \bibinfo{booktitle}{\emph{MM}}. \bibinfo{publisher}{ACM}, \bibinfo{pages}{3070--3078}.
\newblock


\bibitem[Wu et~al\mbox{.}(2021)]%
        {wu2021sgl}
\bibfield{author}{\bibinfo{person}{Jiancan Wu}, \bibinfo{person}{Xiang Wang}, \bibinfo{person}{Fuli Feng}, \bibinfo{person}{Xiangnan He}, \bibinfo{person}{Liang Chen}, \bibinfo{person}{Jianxun Lian}, {and} \bibinfo{person}{Xing Xie}.} \bibinfo{year}{2021}\natexlab{}.
\newblock \showarticletitle{Self-Supervised Graph Learning for Recommendation}. In \bibinfo{booktitle}{\emph{SIGIR}}. \bibinfo{publisher}{ACM}, \bibinfo{pages}{726–735}.
\newblock


\bibitem[Wu et~al\mbox{.}(2022)]%
        {WuLSHGW22}
\bibfield{author}{\bibinfo{person}{Le Wu}, \bibinfo{person}{Junwei Li}, \bibinfo{person}{Peijie Sun}, \bibinfo{person}{Richang Hong}, \bibinfo{person}{Yong Ge}, {and} \bibinfo{person}{Meng Wang}.} \bibinfo{year}{2022}\natexlab{}.
\newblock \showarticletitle{DiffNet++: {A} Neural Influence and Interest Diffusion Network for Social Recommendation}.
\newblock \bibinfo{journal}{\emph{TKDE}} \bibinfo{volume}{34}, \bibinfo{number}{10} (\bibinfo{year}{2022}), \bibinfo{pages}{4753--4766}.
\newblock


\bibitem[Wu et~al\mbox{.}(2019)]%
        {wu2019neural}
\bibfield{author}{\bibinfo{person}{Le Wu}, \bibinfo{person}{Peijie Sun}, \bibinfo{person}{Yanjie Fu}, \bibinfo{person}{Richang Hong}, \bibinfo{person}{Xiting Wang}, {and} \bibinfo{person}{Meng Wang}.} \bibinfo{year}{2019}\natexlab{}.
\newblock \showarticletitle{A neural influence diffusion model for social recommendation}. In \bibinfo{booktitle}{\emph{SIGIR}}. \bibinfo{pages}{235--244}.
\newblock


\bibitem[Wu et~al\mbox{.}(2016)]%
        {wu2016collaborative}
\bibfield{author}{\bibinfo{person}{Yao Wu}, \bibinfo{person}{Christopher DuBois}, \bibinfo{person}{Alice~X Zheng}, {and} \bibinfo{person}{Martin Ester}.} \bibinfo{year}{2016}\natexlab{}.
\newblock \showarticletitle{Collaborative denoising auto-encoders for top-n recommender systems}. In \bibinfo{booktitle}{\emph{WSDM}}. ACM, \bibinfo{pages}{153--162}.
\newblock


\bibitem[Xin et~al\mbox{.}(2023)]%
        {xin2023improving}
\bibfield{author}{\bibinfo{person}{Xin Xin}, \bibinfo{person}{Xiangyuan Liu}, \bibinfo{person}{Hanbing Wang}, \bibinfo{person}{Pengjie Ren}, \bibinfo{person}{Zhumin Chen}, \bibinfo{person}{Jiahuan Lei}, \bibinfo{person}{Xinlei Shi}, \bibinfo{person}{Hengliang Luo}, \bibinfo{person}{Joemon~M Jose}, \bibinfo{person}{Maarten de Rijke}, {et~al\mbox{.}}} \bibinfo{year}{2023}\natexlab{}.
\newblock \showarticletitle{Improving Implicit Feedback-Based Recommendation through Multi-Behavior Alignment}. In \bibinfo{booktitle}{\emph{SIGIR}}. \bibinfo{publisher}{ACM}, \bibinfo{pages}{932--941}.
\newblock


\bibitem[Yang et~al\mbox{.}(2022)]%
        {yang2022diffusion}
\bibfield{author}{\bibinfo{person}{Ling Yang}, \bibinfo{person}{Zhilong Zhang}, \bibinfo{person}{Yang Song}, \bibinfo{person}{Shenda Hong}, \bibinfo{person}{Runsheng Xu}, \bibinfo{person}{Yue Zhao}, \bibinfo{person}{Wentao Zhang}, \bibinfo{person}{Bin Cui}, {and} \bibinfo{person}{Ming-Hsuan Yang}.} \bibinfo{year}{2022}\natexlab{}.
\newblock \showarticletitle{Diffusion models: A comprehensive survey of methods and applications}.
\newblock \bibinfo{journal}{\emph{Comput. Surveys}} (\bibinfo{year}{2022}).
\newblock


\bibitem[Yu and Qin(2020)]%
        {yu2020sampler}
\bibfield{author}{\bibinfo{person}{Wenhui Yu} {and} \bibinfo{person}{Zheng Qin}.} \bibinfo{year}{2020}\natexlab{}.
\newblock \showarticletitle{Sampler design for implicit feedback data by noisy-label robust learning}. In \bibinfo{booktitle}{\emph{SIGIR}}. \bibinfo{publisher}{ACM}, \bibinfo{pages}{861--870}.
\newblock


\bibitem[Yuan et~al\mbox{.}(2023)]%
        {yuan2023interaction}
\bibfield{author}{\bibinfo{person}{Wei Yuan}, \bibinfo{person}{Chaoqun Yang}, \bibinfo{person}{Quoc Viet~Hung Nguyen}, \bibinfo{person}{Lizhen Cui}, \bibinfo{person}{Tieke He}, {and} \bibinfo{person}{Hongzhi Yin}.} \bibinfo{year}{2023}\natexlab{}.
\newblock \showarticletitle{Interaction-level membership inference attack against federated recommender systems}. In \bibinfo{booktitle}{\emph{WWW}}. \bibinfo{publisher}{ACM}, \bibinfo{pages}{1053--1062}.
\newblock


\bibitem[Zhang et~al\mbox{.}(2023a)]%
        {zhang2023denoising}
\bibfield{author}{\bibinfo{person}{Chi Zhang}, \bibinfo{person}{Rui Chen}, \bibinfo{person}{Xiangyu Zhao}, \bibinfo{person}{Qilong Han}, {and} \bibinfo{person}{Li Li}.} \bibinfo{year}{2023}\natexlab{a}.
\newblock \showarticletitle{Denoising and Prompt-Tuning for Multi-Behavior Recommendation}. In \bibinfo{booktitle}{\emph{WWW}}. \bibinfo{publisher}{ACM}, \bibinfo{pages}{1355--1363}.
\newblock


\bibitem[Zhang et~al\mbox{.}(2023b)]%
        {zhang2023sled}
\bibfield{author}{\bibinfo{person}{Shengyu Zhang}, \bibinfo{person}{Tan Jiang}, \bibinfo{person}{Kun Kuang}, \bibinfo{person}{Fuli Feng}, \bibinfo{person}{Jin Yu}, \bibinfo{person}{Jianxin Ma}, \bibinfo{person}{Zhou Zhao}, \bibinfo{person}{Jianke Zhu}, \bibinfo{person}{Hongxia Yang}, \bibinfo{person}{Tat-Seng Chua}, {et~al\mbox{.}}} \bibinfo{year}{2023}\natexlab{b}.
\newblock \showarticletitle{SLED: Structure Learning based Denoising for Recommendation}.
\newblock \bibinfo{journal}{\emph{TOIS}} (\bibinfo{year}{2023}).
\newblock


\bibitem[Zhang et~al\mbox{.}(2017)]%
        {zhang2017autosvd++}
\bibfield{author}{\bibinfo{person}{Shuai Zhang}, \bibinfo{person}{Lina Yao}, {and} \bibinfo{person}{Xiwei Xu}.} \bibinfo{year}{2017}\natexlab{}.
\newblock \showarticletitle{Autosvd++ an efficient hybrid collaborative filtering model via contractive auto-encoders}. In \bibinfo{booktitle}{\emph{SIGIR}}. \bibinfo{publisher}{ACM}, \bibinfo{pages}{957--960}.
\newblock


\bibitem[Zhao et~al\mbox{.}(2023)]%
        {zhao2023popularity}
\bibfield{author}{\bibinfo{person}{Jujia Zhao}, \bibinfo{person}{Wenjie Wang}, \bibinfo{person}{Xinyu Lin}, \bibinfo{person}{Leigang Qu}, \bibinfo{person}{Jizhi Zhang}, {and} \bibinfo{person}{Tat-Seng Chua}.} \bibinfo{year}{2023}\natexlab{}.
\newblock \showarticletitle{Popularity-aware Distributionally Robust Optimization for Recommendation System}. In \bibinfo{booktitle}{\emph{CIKM}}. \bibinfo{publisher}{ACM}, \bibinfo{pages}{4967--4973}.
\newblock


\end{thebibliography}
}

\end{document}